\documentclass[fleqn]{aa}

\usepackage{amsmath}
\usepackage{amssymb}
\usepackage{txfonts}

\usepackage{time}
\usepackage[clockwise]{rotating}
\usepackage[dvipsnames]{xcolor}
\usepackage{subfig}

\usepackage{url}

\usepackage{natbib}
\bibpunct{(}{)}{;}{a}{}{,}

\usepackage{epic}
\usepackage{eepic}

\usepackage{overpic}
\usepackage{multirow}
\usepackage{balance}

\def\abs#1{|#1|}

\DeclareMathOperator{\jinc}{jinc}
\def\Tp{T_\text{p}}

\def\Tf{T_\text{f}}

\def\tp{\tau_\text{p}}
\def\tf{\tau_\text{f}}

\def\Fourier{\operatorname{\mathfrak{F}}}

\def\xp{x_\mathrm{p}}
\def\yp{y_\mathrm{p}}
\def\xf{x_\mathrm{f}}
\def\yf{y_\mathrm{f}}

\def\E{\mathbf{E}}
\def\S{\mathbf{S}}
\def\H{\mathbf{H}}

\def\Stilt{\ensuremath{S_\text{tilt}}}
\def\Satm{\ensuremath{S_\text{atm}}}
\def\Swf{\ensuremath{S_\text{wf}}}

\def\Smono{\ensuremath{S_\text{mono}}}

\begin{document}

\date{Draft \now\ \today}
\frenchspacing


\title{A tilted interference filter in a converging beam}

\author{Mats G. L{\"o}fdahl\inst{1,2} \and Vasco M. J.
  Henriques\inst{1,2} \and Dan Kiselman\inst{1,2}}

\offprints{M. L{\"o}fdahl, \email{mats@astro.su.se}.}

\institute{Institute for Solar Physics, Royal Swedish Academy of Sciences,
AlbaNova University Center, 106\,91 Stockholm, Sweden \and
Stockholm Observatory, Dept. of Astronomy, Stockholm University,
AlbaNova University Center, 106\,91 Stockholm, Sweden}


\abstract{Narrow-band interference filters can be tuned toward shorter
  wavelengths by tilting them from the perpendicular to the optical
  axis. This can be used as a cheap alternative to real tunable
  filters, such as Fabry-P\'erot interferometers and Lyot filters. At
  the Swedish 1-meter Solar Telescope, such a setup is used to scan
  through the blue wing of the \ion{Ca}{ii}~H line. Because the filter
  is mounted in a converging beam, the incident angle varies over the
  pupil, which causes a variation of the transmission over the pupil,
  different for each wavelength within the passband. This causes
  broadening of the filter transmission profile and degradation of the
  image quality.}
{We want to characterize the properties of our filter, at normal
  incidence as well as at different tilt angles. Knowing the broadened
  profile is important for the interpretation of the solar images.
  Compensating the images for the degrading effects will improve the
  resolution and remove one source of image contrast degradation. In
  particular, we need to solve the latter problem for images that are
  also compensated for blurring caused by atmospheric turbulence.}
{We simulate the process of image formation through a tilted
  interference filter in order to understand the effects.  We test the
  hypothesis that they are separable from the effects of wavefront
  aberrations for the purpose of image deconvolution. We measure the
  filter transmission profile and the degrading PSF from calibration
  data.}
{We find that the filter transmission profile differs significantly
  from the specifications. We demonstrate how to compensate for the
  image-degrading effects. Because the filter tilt effects indeed
  appear to be separable from wavefront aberrations in a useful way,
  this can be done in a final deconvolution, after standard image
  restoration with Multi-Frame Blind Deconvolution/Phase Diversity
  based methods. We illustrate the technique with real data.}
{}
\keywords{Instrumentation: interferometers -- Methods: observational
  -- Techniques: image processing -- Techniques: imaging spectroscopy}

\maketitle 


\section{Introduction}
\label{sec:intro}  

Narrowband (NB) interference filters can be tuned in wavelength by
varying the incident angle. This is often used to allow for filter
manufacturing margins by specifying the peak to the red of the wanted
wavelength and then mounting the filter at a small angle to
compensate. At the Swedish 1-m Solar Telescope
\citep[SST;][]{scharmer03new}, the effect is used for scanning through
the blue wing of the \ion{Ca}{ii}~H line by use of a narrow-band (NB)
filter mounted on a motor-controlled rotation stage. The tuning is
possible because the filter cavity is geometrically wider for a ray
intersecting with the filter at an angle from normal incidence.

However, even when the filter is mounted with no tilt with respect to
the optical axis, rays arrive at a variety of angles. This has
consequences for the spectral resolution as well as for the spatial
resolution \citep{beckers98effect}. Depending on the optical
configuration, the spread in incident angles may cause a broadening of
the transmission profile or variations in central wavelength over the
field of view (FOV) or both. A problem related to the broadening is
that the resolution suffers in telescopes that operate near the
diffraction limit. The amplitude of the pupil transmission
coefficient, as seen through the filter, is apodized by an amount that
varies with the wavelength. Beckers evaluated the resolution effects
of a Fabry-P\'erot interferometer (FPI) used in the pupil plane of a
telecentric setup. \Citet{luhe00high} included also the phase of the
field transmission in their analysis. \citet{scharmer06comments}
showed that the dominating part of the phase effects is in the form of
a wavelength independent, quadratic pupil phase that can be
compensated with a simple re-focusing of the detector.

The effects investigated by the authors cited above are greater the
narrower the bandpass, making this a problem primarily for FPIs with
FWHMs in the picometer range. However, as we will demonstrate in this
paper, there are significant effects also for filters with a FWHM of
$\sim$0.1~nm when tilted by a few degrees. We analyze a NB
double-cavity filter, mounted in a converging beam at an angle from
the optical axis. We show that the resulting degradation of image
quality can be compensated for by deconvolution and that images
degraded by this effect as well as atmospheric seeing can be
significantly improved. The effect is apparently separable from phase
aberrations in the sense that they can be modeled as separate transfer
functions, to be applied in sequence. This is beneficial for wavefront
sensing with the Multi-Frame Blind Deconvolution/Joint Phase Diverse
Speckle (MFBD/JPDS) formalism of \citet{lofdahl02multi-frame}.

We illustrate the technique with images from the SST recorded through
a \ion{Ca}{ii}~H filter mounted on a computer controlled rotating
stage. We use simulations of such data to illustrate the separability
of the apodisation effects and phase aberrations.  Appropriate and
sufficient information needed to compensate for the image degradation
can be obtained by calibration using pinhole images.

\section{Theory and notation}
\label{sec:theory}

\subsection{Tilted cavities}
\label{sec:tilting-filter}

In order to understand what happens when we tilt a filter, we begin by
deriving the transmission coefficient for the electric field. The
matrix formalism described by \citet[Sect. 5.4]{klein86optics}
implements multiple reflections and transmissions in both directions
through many different layers. Assume light incident on a stack of $N$
layers from the left. We define the column vectors
\begin{equation}
  \label{eq:4}
  \E_j = 
  \left[
    \begin{array}{c}
      E_{\text{L}j}\\
      E_{\text{R}j}
    \end{array}
    \right]
\qquad\text{and}\qquad
  \E'_j = 
  \left[
    \begin{array}{c}
      E'_{\text{L}j}\\
      E'_{\text{R}j}
    \end{array}
    \right],
\end{equation}
where $\E_j$ represents the electric field on the left side of layer
$j$ and $\E'_j$ the field on the right side of the same layer. Indices
L and R denote field moving to the left and right, respectively.

The relationship between the field in the incident medium and the
field in the final medium can then be written as
\begin{equation}
  \label{eq:7}
  \E'_1 = \S_{1N}\E_N,
\end{equation}
with the stack matrix 
\begin{equation}
  \label{eq:8}
  \S_{1N} = \left[
    \begin{array}{cc}
      S_{11} &  S_{12}\\
      S_{21} &  S_{22}\\
    \end{array}
  \right] = \H_{12}\mathbf{L}_{2}\cdots\mathbf{L}_{N-1}\H_{N-1,N}.
\end{equation}
The interface transition matrix $\H_{ij}$ and the layer matrix $\mathbf{L}_j$
are defined as
\begin{equation}
  \label{eq:9}
  \H_{ij} = \frac{1}{\tau_{ij}} \left[
    \begin{array}{cc}
      1& \rho_{ij} \\
      \rho_{ij} & 1 \\
    \end{array}
  \right]
  \qquad\text{and}\qquad  
  \mathbf{L}_i =  \left[
    \begin{array}{cc}
      e^{-\text{i}\beta_j} & 0 \\
      0 & e^{\text{i}\beta_j} \\
    \end{array}
  \right],
\end{equation}
repectively, where $\tau_{ij}$ and $\rho_{ij}$ are the transmission
and reflection coefficients, respectively, of the interface between
medium $i$ and medium $j$, and
\begin{equation}
  \beta_j = \frac{4\pi}{\lambda}n_jh_j\cos\alpha_j .
  \label{eq:2}
\end{equation}
Here $n_j$ is the refractive index of medium $j$, $h_j$ is the
geometrical thickness of the cavity, and $\alpha_j$ is the angle of
the ray from the interface surface normal within medium $j$.  Using
these definitions and the boundary condition $E_{\text{L}N}=0$ (in the
final medium there is no field moving to the left), the transmission
coefficient of the stack can be written as
\begin{equation}
  \label{eq:11}
  \tau = \frac{E_{\text{R}N}}{E'_{\text{R}1}} = \frac{1}{S_{22}}.
\end{equation}

Applying this formalism to a single cavity with index $n_2$ between
media of index $n_1=n_N$, we get
\begin{equation}
  \label{eq:12}
  \tau_\text{sc} =
  \frac{\tau_{12}\tau_{21}e^{-\text{i}\beta_2}}{1-\rho^2_{12}e^{-\text{i}2\beta_2}}.
\end{equation}
If $n_2$, $h_2$, $d_2$, $\tau_{12}$, $\tau_{21}$, $\rho_{12}$, and
$\rho_{12}$ are known, like for a FPI, this expression can be used to
calculate the transmission of the cavity for any incident angle.
However, manufacturers of dielectric filters usually only disclose the
intensity transmittance ($T=\abs{\tau}^2$) at normal incidence.

The transmission coefficient is a function of the wavelength as well
as the incident angle through $\beta_2$. When the angle changes, the
wavelength needed to keep $\beta_2$ (and thereby $\tau$) constant also
changes. The effect of changing the angle is therefore to shift the
transmission profile in wavelength.
With the assumption of small angles, repeated application of Snell's
law, and trivial algebra, the well known \citep[e.g.,][chapter
7]{smith90modern} expression
\begin{equation}
  \lambda' = 
   \lambda\sqrt{1 - \left(\frac{n_1}{n_2}\sin\alpha\right)^2}, 
  \label{eq:lambda_a}
\end{equation}
can be derived, where $\lambda'$ is the shifted wavelength and we
define $\alpha=\alpha_1$.

With double cavities we get
\begin{equation}
  \label{eq:13b}
  \tau_\text{dc} =  \frac{\tau_{12}\tau_{23}\tau_{34} \cdot e^{-\text{i}\overline\beta}}
  {1 + \overline\rho^2 e^{-\text{i}2\overline\beta} - \rho_{23}\rho_{34}e^{-\text{i}2\beta_2}-\rho_{12}\rho_{23}e^{-\text{i}2\beta_3}},
\end{equation}
where we have defined $\overline\beta=(\beta_2+\beta_3)$, and
$\overline\rho^2=\rho_{12}\rho_{23}+\rho_{23}\rho_{34}+\rho_{12}\rho_{34}$.
The numerator shifts as given by the shift of $\overline\beta$, which
means it behaves like the transmission profile for a single cavity and
an ``effective'' refractive index $n$ can be defined. The last two
terms in the denominator shift differently but the behavior of the
numerator should dominate.

\subsection{Pupil transmission in a converging beam}
\label{sec:pupil-funct}

For a piece of optics in a converging beam, the incident angle of a
light ray varies with the orientation of the optics relative to the
optical axis, but also with the location within the pupil that the ray
emanates from. As we have seen, the transmission coefficient
$\tau(\lambda)$ of an interference filter varies with the incident
angle, so that it is shifted toward shorter wavelengths.

For incoming light with a wavelength $\lambda$, the generalized pupil
function may be written as
\begin{equation}
  \label{eq:pupil1}
  P(\xp,\yp,\lambda;\phi,\theta) = 
  \tp(\xp,\yp,\lambda;\theta)\, A(\xp,\yp)\exp\{\mathrm{i}\phi(\xp,\yp)\},
\end{equation}
where $(\xp,\yp)$ are coordinates within the pupil (here normalized to
unit diameter), $A$ is a binary function representing the geometrical
shape of the pupil stop, $\mathrm{i}$ is the imaginary unit, $\theta$
is the tilt angle of the filter around the $y$ axis, and $\phi$ is the
phase aberrations from the atmosphere, the telescope, and any other
optics, independent of $\lambda$.
The filter's complex transmission coefficient as seen through the
pupil can be written as
\begin{equation}
  \label{eq:1}
  \tp(\xp,\yp,\lambda;\theta) =
  \tf\bigl(\lambda'(\lambda,\alpha(\xp,\yp;\theta))\bigr),
\end{equation} 
where $\tf$ is the normal incidence complex transmission coefficient
of the electric field vector through the filter, and $\lambda'$ is the
shifted wavelength given by Eq.~(\ref{eq:lambda_a}). In Appendix
\ref{sec:incident-angle} we derive an expression for the incident
angle, $\alpha$, on a tilted filter in a converging beam. With the
added assumption that the detector is small, we can write it as
\begin{equation}
  \alpha(\xp,\yp; \theta) 
  = 
  \arccos\left(
    \frac{F\cos\theta + \xp\sin\theta}
         {(F^2 + \xp^{2} + \yp^{2})^{1/2}}
  \right),
  \label{eq:alpha}
\end{equation}
where $F$ is the effective F-ratio.

\subsection{Point spread function and optical transfer function}
\label{sec:optic-transf-funct}

Because of Eq.~(\ref{eq:lambda_a}), we know that NB interference
filters can be tuned by tilting the filter normal from the optical
axis. In addition to the desired change in peak wavelength, the
tilting has consequences for the width of the profile and for the
shape of the PSF because of Eq.~(\ref{eq:alpha}).

We model the image formation process as a space-invariant (valid for
sufficiently small sub-fields) convolution between an unknown object
$f$ and a point spread function $s_k$, the Fourier transforms of which
are $F$ and $S_k$, where the index $k$ corresponds to a particular
data frame. An additive Gaussian noise term $n_k$ is assumed. The
Fourier transform $D_k$ of the observed image $d_k$ is then related to
$F$, $S_k$ and $N_k$ via
\begin{equation}
  D_k = F S_k + N_k .
\label{eq:15}
\end{equation}
These quantities are all functions of the spatial frequency
coordinates $u$ and $v$ but to ease the notation, we will not write
them out.  The assumption of additive Gaussian noise leads to a simple
Maximum Likelihood expression for an estimate of the object through
multi-frame deconvolution,
\begin{equation}
  \label{eq:21}
  \hat F =  H \sum_k D_k S_k^* \biggm/ \sum_k\abs{S_k}^2 ,
\end{equation}
where $H$ is a low-pass noise filter and $*$ used as a superscript
denotes the complex conjugate.

Because the profiles are shifted by different amounts over the pupil,
the transmittance of light of different wavelengths through the pupil
is not uniform. At every wavelength within the broadened passband, the
transmittance of light from different parts of the pupil corresponds
to different parts of the normal-incidence filter profile. Because
light interferes only with light of the same wavelength, $S_k$ cannot
be calculated with the usual quasi-monochromatic assumption. Instead,
it must be modeled as the integrated, monochromatic contributions from
the different wavelengths within the passband. With a discrete
approximation, we can write
\begin{equation}
  \label{eq:Sk}
  S_k = S(\phi_k,\theta) = 
  \delta\lambda \sum_\ell^L 
  \Fourier\left\{ 
    \left| 
      \Fourier^{-1} P(\xp,\yp,\lambda_\ell;\phi_k,\theta)
    \right|^2
  \right\},
\end{equation} 
where the summation is performed over $L$ samples distributed over the
broadened passband.

What we would really want is for $S_k$ to be separable in two parts:
1) the usual OTF, including the pupil geometry, the atmospheric
wavefront, as well as any other wavefront components introduced in the
optics, and 2) another transfer function, compensating for the filter
tilt effects. We would like to write
\begin{equation}
  \label{eq:18}
  S(\phi_k,\theta) \approx \Swf(\phi_k) \cdot \Stilt(\theta),
\end{equation}
where
\begin{equation}
  \label{eq:20}
  \Swf(\phi) =  
  \Fourier\left\{ 
     \left| 
      \Fourier^{-1} A(\xp,\yp) \exp\{\mathrm{i}\phi(\xp,\yp)\}
    \right|^2
  \right\} 
\end{equation}
and
\begin{equation}
  \label{eq:19}
  \Stilt(\theta) = \frac{1}{S_0}
      \delta\lambda \sum_\ell^L  \Smono(\lambda_\ell,\theta),
\end{equation}
where we define the monochromatic tilt OTF as
\begin{equation}
  \label{eq:14}
  \Smono(\lambda,\theta) = 
  \Fourier\left\{ \left| \Fourier^{-1} P(\xp,\yp;\lambda,0,\theta)\right|^2
  \right\}.
\end{equation}
In the expression for \Stilt{}, we choose to include the pupil
geometry and then divide it away as the diffraction limited MTF,
$S_0=\Swf(0)$. This way it becomes natural not to define $\tp$ outside
of the pupil. For simplicity of notation, and for the purpose of this
discussion, we incorporate also any intentional phase diversity (PD)
focus contribution in $\phi_k$.

We explore this separability in Sect.~\ref{sec:point-spre-funct}
below.  The advantages with separability, for our purposes, is that
the correction can be applied after restoration of the images for
atmospheric turbulence effects, i.e., MFBD-based methods or speckle
interferometry. And, since we assume the variation over the field of
view is negligible, it can be done for the entire FOV in a single
operation instead of by subfields like for the anisoplanatic
atmosphere.

\section{A 0.1~nm wide Ca~\textsc{ii}~H 396.9~nm filter}
\label{sec:ca-textscii-h}

\subsection{Filter properties}
\label{sec:filt-prop}

\begin{figure}[!t]
  \centering
  \includegraphics[bb=132 201 524 633,width=0.9\linewidth]{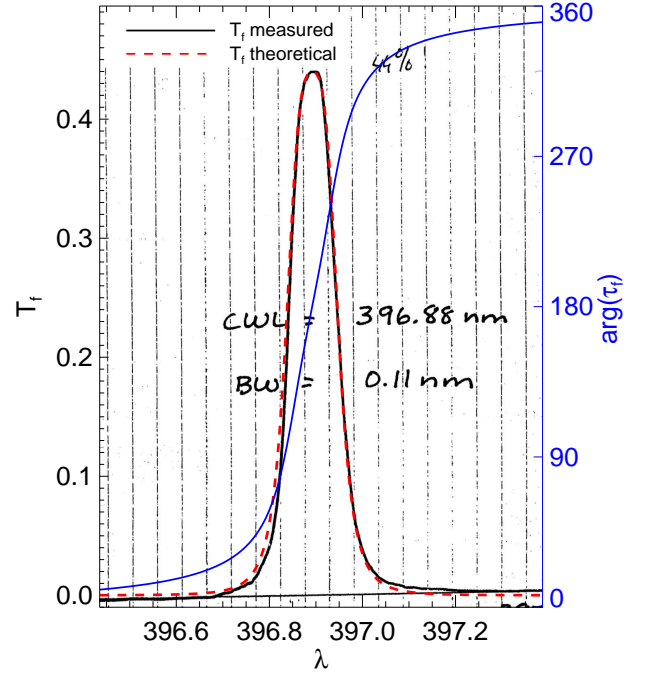}
  \caption{Filter profiles for normal incidence on the \ion{Ca}{ii}~H
    filter. Solid black: Transmittance, $\Tf(\lambda)$, measured by
    the manufacturer before delivery of the filter; Dashed red:
    theoretical transmittance; Blue: theoretical transmission phase,
    $\arg(\tf(\lambda))$.}
  \label{fig:barr_profiles}
\end{figure}

As the specific case, motivating this research, we discuss a
\ion{Ca}{ii}~H filter, regularly used at the SST.  The filter,
manufactured by Barr Associates, Inc., was delivered in August 2000.
The coating parameters are not public so we cannot calculate $\tf$ but
along with the filter we got the measurements shown in
Fig.~\ref{fig:barr_profiles}. The central wavelength was specified to
$\lambda_\mathrm{c}=396.88$~nm, the peak transmittance, $\max\Tf$, to
44\%, and the FWHM to 0.11~nm (at $23\degr$C and normal incidence).

In order to use Eq.~(\ref{eq:lambda_a}), we need the effective
refractive index. \citet[page~26]{rouppe02sunspot} contacted the
manufacturer after the filter was delivered and got the value
$n=1.596$. Because we wanted to explore the effects of the phase,
\citet{potter04private} later provided some design data: the phase of
the filter transmission coefficient, $\arg\{\tf(\lambda)\}$, as well
as the curve form of the transmittance profile in digital form, see
the red and blue curves in Fig.~\ref{fig:barr_profiles}. The
wavelength range for both quantities is $394.9\text{
  nm}\le\lambda\le398.9\text{ nm}$. We show here the central part,
rescaled to the measured peak transmittance of 44\%.

The light beam from the telescope forms a pupil image, where our
adaptive optics \citep{scharmer03adaptive} deformable mirror is
mounted. From there, the beam is re-imaged by a field lens with a
focal length of $\sim$1.5~m, making a F/46 beam. As seen from the
pupil image, through the re-imaging lens, the 7~mm center to corner
distance of our 10~mm by 10~mm detector corresponds to an angular size
of $0.007/1.5$~rad or 0\fdg27. As we will see, angles of this size has
negligible impact so we are justified in ignoring variations over the
FOV.
The image scale, 0\farcs034/pixel, made by the re-imaging lens was
established by observations of the Venus transit in 2004
\citep{kiselman08solar}.

\begin{figure}[!t]
  \centering
  \subfloat[\label{fig:barr_transmissionx}]{\includegraphics[bb=34 60 430 272,clip,width=0.95\linewidth]{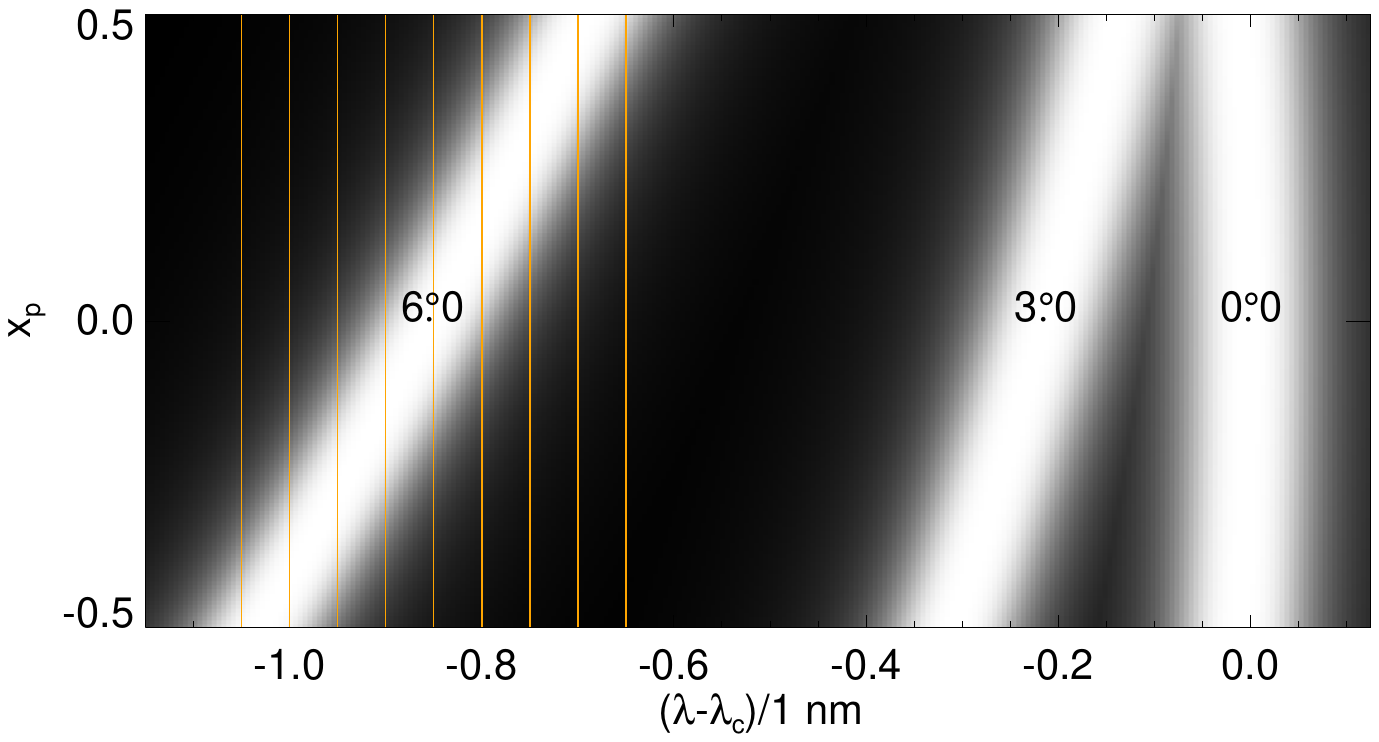}}\\
  \subfloat[\label{fig:barr_transmissionP}]{\includegraphics[width=0.95\linewidth]{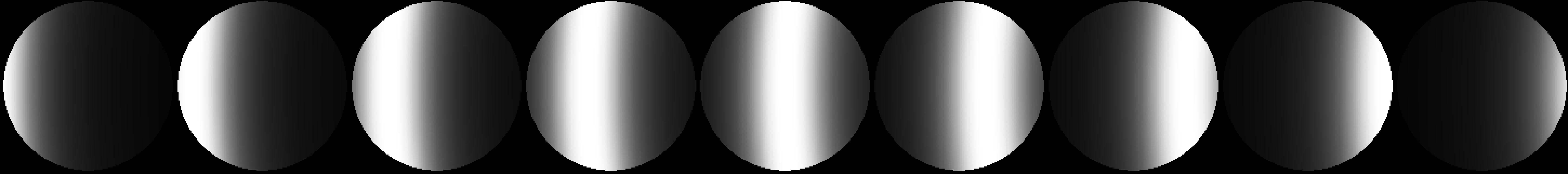}}\\
  \subfloat[\label{fig:barr_phaseP}]{\includegraphics[width=0.95\linewidth]{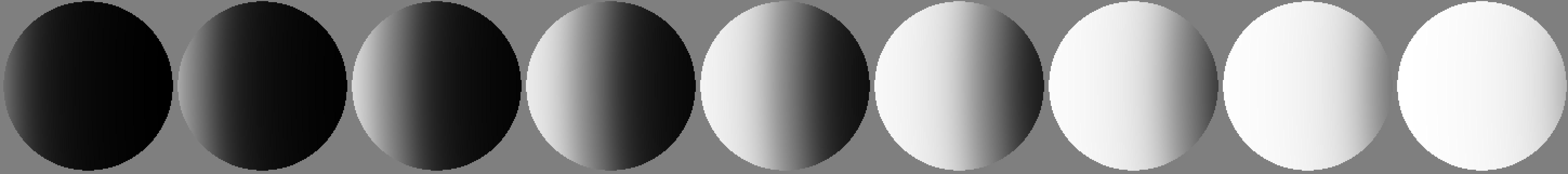}}
  \caption{The transmission coefficient, $\tp$, profile data cube,
    calculated from the manufacturer's theoretical profiles by use of
    Eqs.~(\ref{eq:pupil1})--(\ref{eq:alpha}).
    \textbf{\protect\subref{fig:barr_transmissionx}} The modulus for
    $\yp=0$, $\abs{\tp(\xp,0,\lambda;\theta)}$, superimposed for
    $\theta=0\degr$, 3\degr{}, and 6\degr{} as labeled.
    \textbf{\protect\subref{fig:barr_transmissionP}}~The pupil
    apodisations, $\abs{\tp(\xp,\yp,\lambda_\ell;6\degr)}$, sampled at
    the wavelengths $\lambda_\ell$ indicated with the orange lines
    in~\protect\subref{fig:barr_transmissionx}.
    \textbf{\protect\subref{fig:barr_phaseP}}~The corresponding pupil
    phases, $\arg\{\tp(\xp,\yp,\lambda_\ell;6\degr)\}$.}
  \label{fig:barr_transmission_cube}
\end{figure}

The pupil apodisations vary with tilt angle, which is illustrated in
Fig.~\ref{fig:barr_transmission_cube}. Increasing the angle moves the
passband from the \ion{Ca}{ii}~H line core (formed in the solar
chromosphere) toward the blue wing (formed in the photosphere). In the
axis corresponding to the tilt direction, the transmittance profile
peak position varies over the pupil
(Fig.~\ref{fig:barr_transmissionx}).  In the perpendicular direction, 
the peak position varies only slightly with distance from the center
of the pupil. Sample pupil apodisations for one of the tilt angles are
shown in Fig.~\ref{fig:barr_transmissionP}. Because of the angles of
the transmittance ridges in Fig.~\ref{fig:barr_transmissionx}, larger
tilt angles result in narrower apodisations over the pupil for each
wavelength. For 0\fdg0, the transmittance varies with wavelength, but
with small variations within the pupil. For larger tilt angles, the
apodisation function looks more and more like a band sweeping across
the pupil.

The transmittance phase, $\arg(\tf)$, has an approximately linear
gradient where the modulus, $\abs{\tf}$, has its peak, see
Fig.~\ref{fig:barr_profiles}. The phase profiles are shifted the same
way as the transmittance profiles, so the phase gradient peaks at the
same location in the pupil as the transmittance, see
Figs.~\ref{fig:barr_transmissionP} and \ref{fig:barr_phaseP}.

\subsection{Profiles}
\label{sec:profiles}

\begin{figure}[!t]
  \includegraphics[bb=50 12 483 338,clip,width=\linewidth]{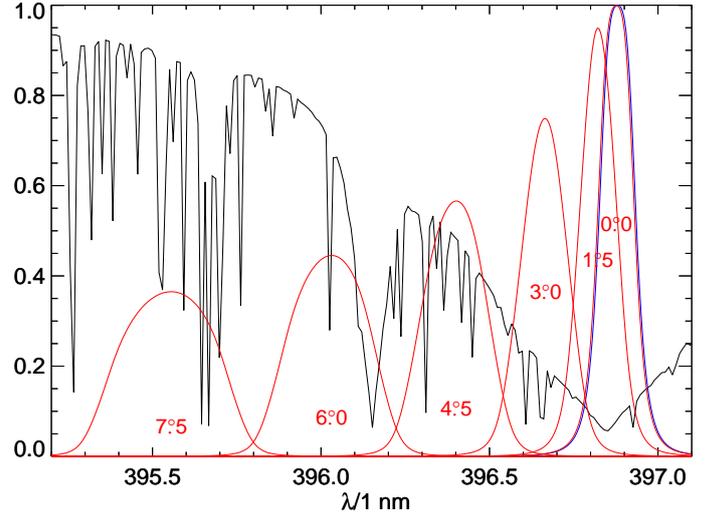}
  \caption{Integrated filter transmittance profiles,
    $\sum_{x_p,y_p}\Tp(\lambda,\theta)$ (red) corresponding to
    filter tilt angles ($\theta$) as indicated in the figure. The
    $\theta=0\degr$ profile is virtually identical to the normal
    incidence filter profile, $\Tf(\lambda)$ (blue). The black line is
    the \ion{Ca}{ii}~H spectrum of the synthetic data of
    Sect.~\ref{sec:synthetic-data}.}
  \label{fig:profiles}
\end{figure}

Integrating over the pupil results in broadened profiles, as noted by
\citet{beckers98effect}, because of the varying shifts in the $\xp$
direction. For our filter, $\theta=0\fdg0$ results in a transmission
profile that is virtually identical to the normal incidence profile
but for angles of a few degrees, the broadening is significant, see
the broadening as calculated for the manufacturer's profile in
Fig.~\ref{fig:profiles}.

In order to test the predicted broadening and wavelength shifts, the
filter was examined using the TRI-Port Polarimetric Echelle-Littrow
spectrograph (Kiselman et al. in prep.) at the SST on 2008-05-13. The
filter on its rotatable mount was placed in front of the spectrograph
slit and the telescope was pointed to solar disk center. Spectra (2000
wavelength points in the range
$396.1\text{~nm}\le\lambda\le397.1\text{~nm}$) were collected without
the filter, as well as with the filter at 18 different tilt angles
that turned out to be $-2\fdg16\le\theta\le5\fdg84$, i.e., angles on
both sides of the symmetry point are well represented.
The spectra were reduced with standard methods. Filter profiles were
then calculated by normalizing the filtered spectra with the
unfiltered spectrum. The resulting profiles contain small residues
from blending spectral lines which have not been completely removed by
the normalization. This is due to the presence of straylight within
the spectrograph, which is corrected for in the reductions but the
procedure obviously does not work perfectly for spectra of this kind.
We deem the resulting profiles good enough for our current purpose,
however.

The wavelength scale was calibrated with the help of a spectral atlas
\citep{brault87spectral} and transformed to the laboratory rest frame.
The zero angle is then given by symmetry and the peak wavelength for
normal incidence, $\lambda_\text{c}=396.89$~nm, follows.

\begin{figure}[!t]
  \centering
  \def\tilewidth{\linewidth}
  \includegraphics[bb=30 11 495 336,width=\tilewidth]{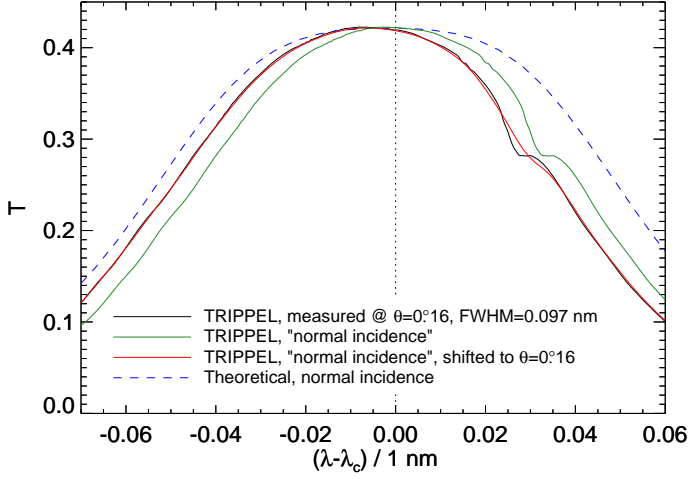}
  \caption{The filter normal incidence transmittance as measured with
    TRIPPEL. Green: the measured profile shifted towards the red to
    compensate for tilt angle, our measured $\Tf$. Red: model of black
    curve using measured $\Tf$. Black: measured profile with smallest
    angle. Blue dashed: theoretical $\Tf$ (peak reduced to 42\%).}
  \label{fig:trippel-profiles-near-normal}
\end{figure}
\begin{figure}[!t]
  \centering
  \def\tilewidth{\linewidth}
  \includegraphics[bb=30 11 495 336,width=\tilewidth]{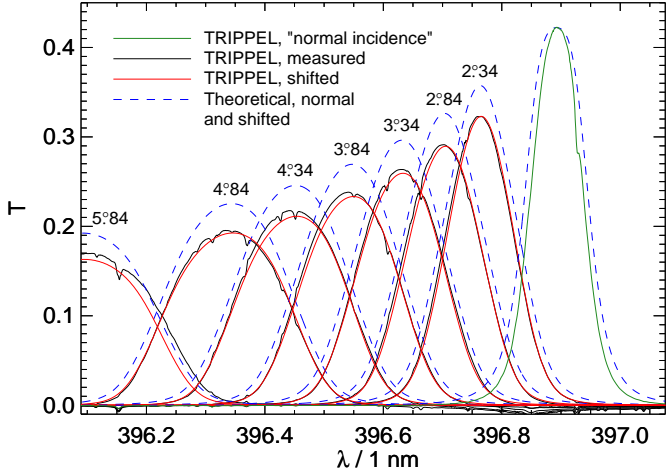}
  \caption{TRIPPEL transmittance measurements at $\theta>2\degr$.
    Black: measured profiles. Red: models of measured curves using the
    TRIPPEL-measured ``normal incidence'' profile. Blue dashed: models
    of measured curves using the theoretical normal incidence profile
    with peak reduced to 42\%.}
  \label{fig:trippel-profiles-all}
\end{figure}

In Fig.~\ref{fig:trippel-profiles-near-normal}, we show the
transmittance profile measured with the smallest angle, 0\fdg16. For
such small angles, while tilting shifts the central wavelength, it
does not broaden the profile significantly. Therefore this should be a
good measure of the normal incidence profile $\Tf$. We demonstrate
that this is so by relocating the measured profile to the symmetry
point. Using this to model the measured curve with the filter tilt and
pupil apodisation formalism, it is clear that this is just the amount
needed for the tilt to shift it back to the measured position and that
the match in profile shape is excellent. The only thing that happens
to the profile shape is that the small artifact in the red flank is
smoothed out slightly. The FWHM of the measured $\Tf$ is 0.097~nm
rather than the 0.11~nm value from the manufacturer, a significant
difference of more than 10\%.

The consequences for large angles are shown in
Fig.~\ref{fig:trippel-profiles-all}. Using the TRIPPEL-measured
normal-incidence profile, we can model the measurements very well,
which means the profile shift and broadening based on single-cavity
theory work well and that the effective refractive index, $n=1.596$,
must be a good approximation. Shifting the manufacturer's profile
produces profiles that broaden in the right way, but where the peak
transmittance does not decrease with tilt angle fast enough. This is
caused by the wider normal incidence profile and the fact that the
broadening preserves the area under the profiles.

If a manufacturer is not able or willing to provide an effective
refractive index, it should be possible to infer a useful estimate
from data like these, by fitting Eq.~(\ref{eq:lambda_a}) to the
peaks of the shifted profiles.
 
\subsection{Point spread functions and Strehl ratios}
\label{sec:point-spre-funct}

The pupil phase for any particular $\theta$ varies with $\lambda$, so
it is not obvious what phase, if any, would be effective as a
correction of the summed \Stilt. In this section we investigate this.
We use the theoretical $\tf$ profile here, because we have modulus and
phase that go together.

\begin{figure}[!t]
  \def\figdir{/home/mats/data/pupil-apodisation/PSFs}
  \def\tilewidth{0.15\linewidth}
  \subfloat[0\fdg0]{
    \begin{minipage}[c]{\tilewidth}
      \includegraphics[width=\linewidth]{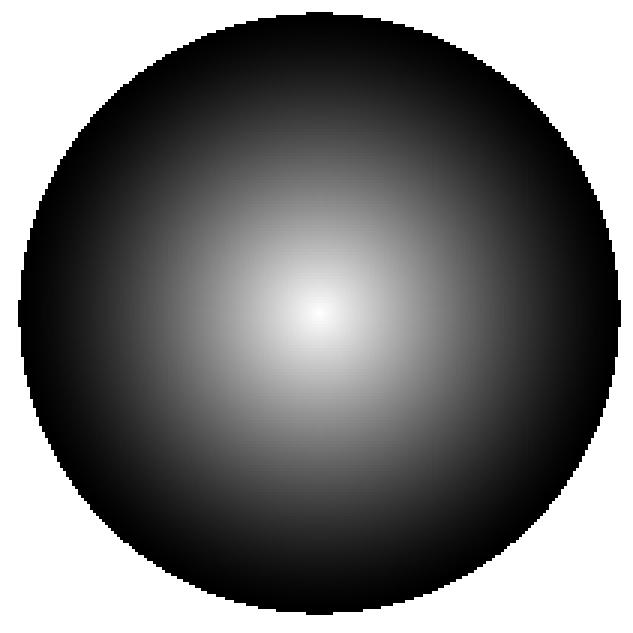}\\
      \includegraphics[width=\linewidth]{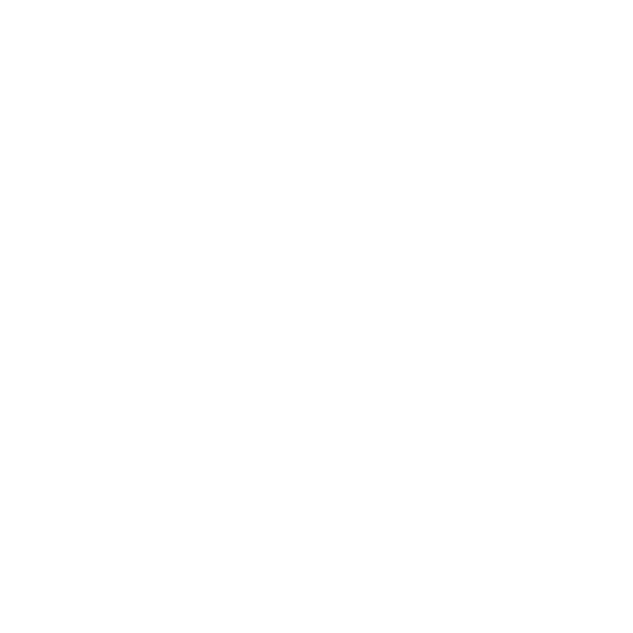}
    \end{minipage}}     
  \subfloat[1\fdg5]{
    \begin{minipage}[c]{\tilewidth}
      \includegraphics[width=\linewidth]{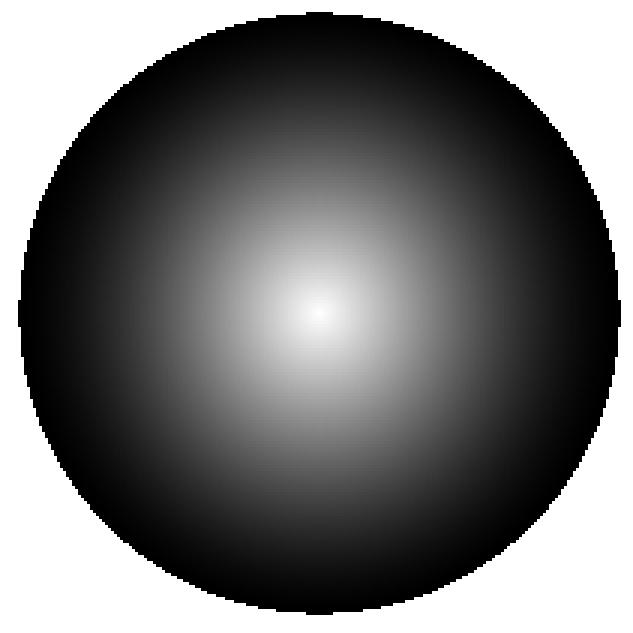}\\
      \includegraphics[width=\linewidth]{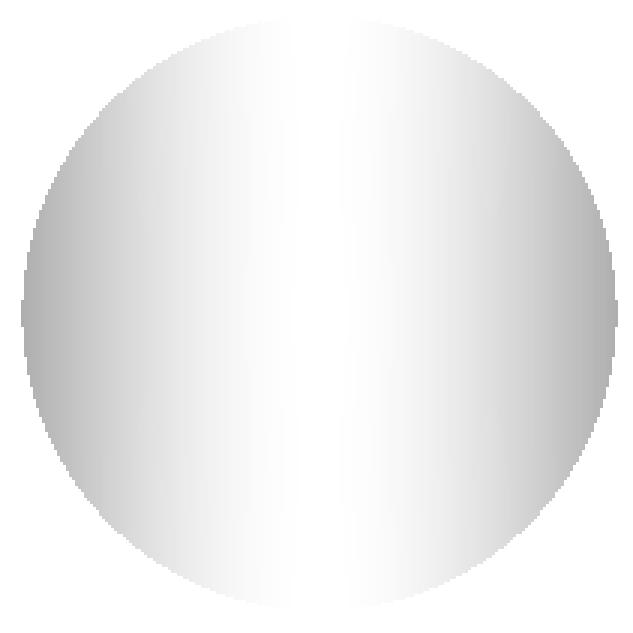}
    \end{minipage}}     
  \subfloat[3\fdg0]{
    \begin{minipage}[c]{\tilewidth}
      \includegraphics[width=\linewidth]{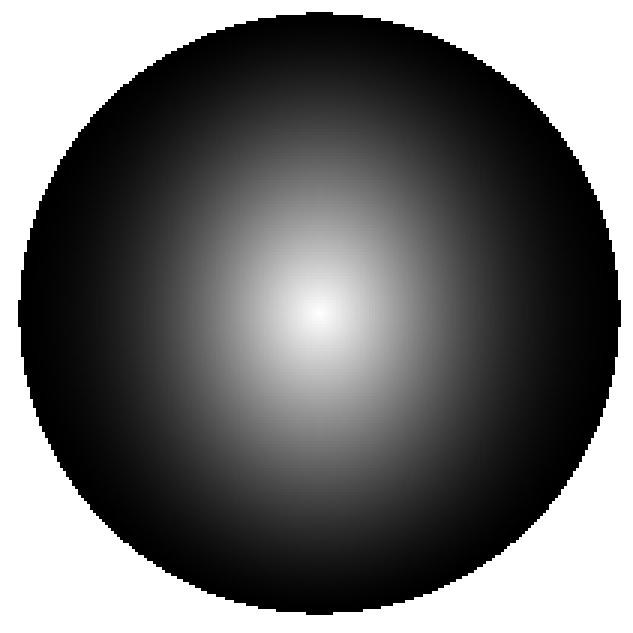}\\
      \includegraphics[width=\linewidth]{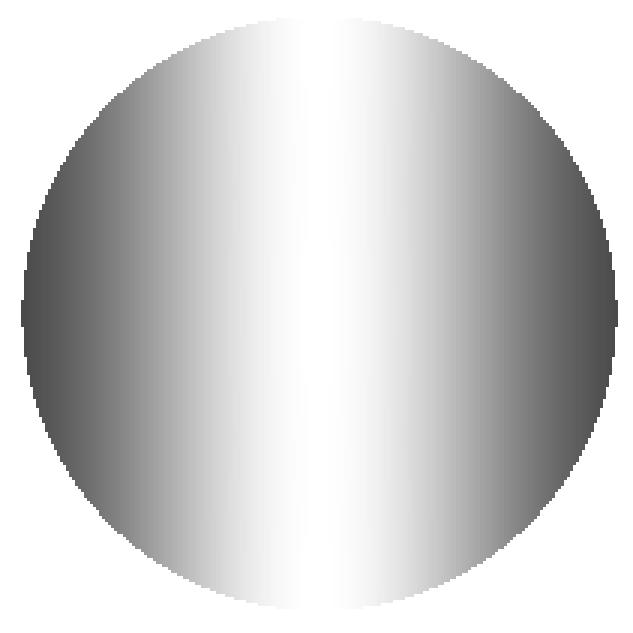}
    \end{minipage}}     
  \subfloat[4\fdg5]{
    \begin{minipage}[c]{\tilewidth}
      \includegraphics[width=\linewidth]{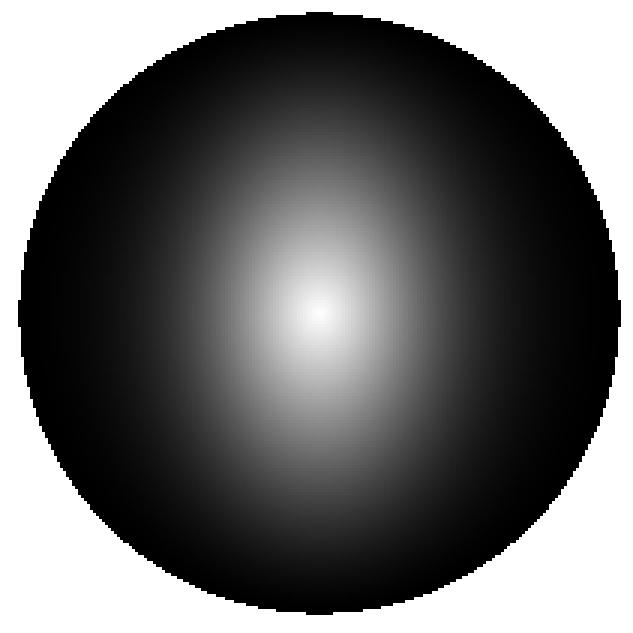}\\
      \includegraphics[width=\linewidth]{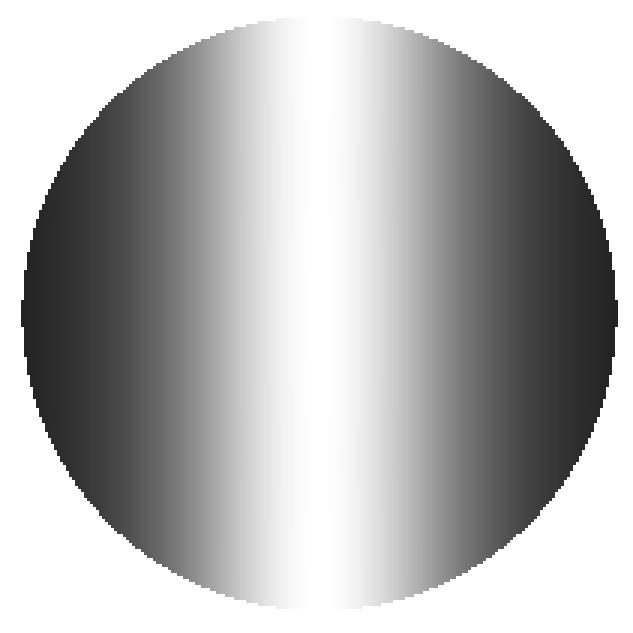}
    \end{minipage}}     
  \subfloat[6\fdg0]{
    \begin{minipage}[c]{\tilewidth}
      \includegraphics[width=\linewidth]{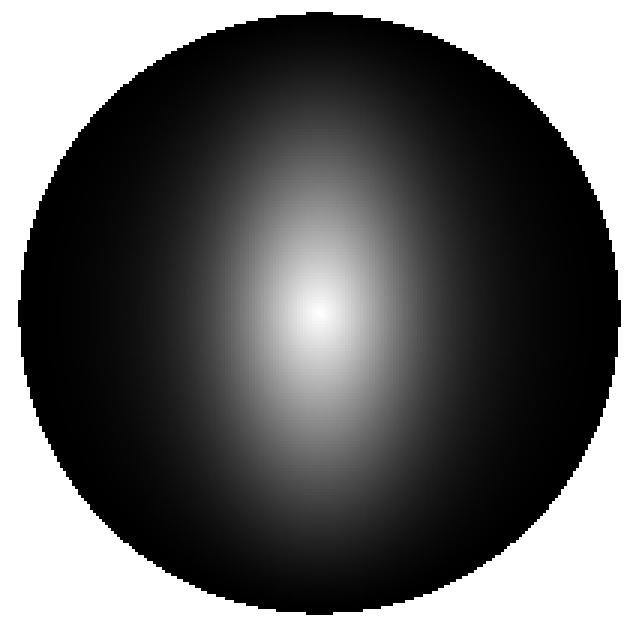}\\
      \includegraphics[width=\linewidth]{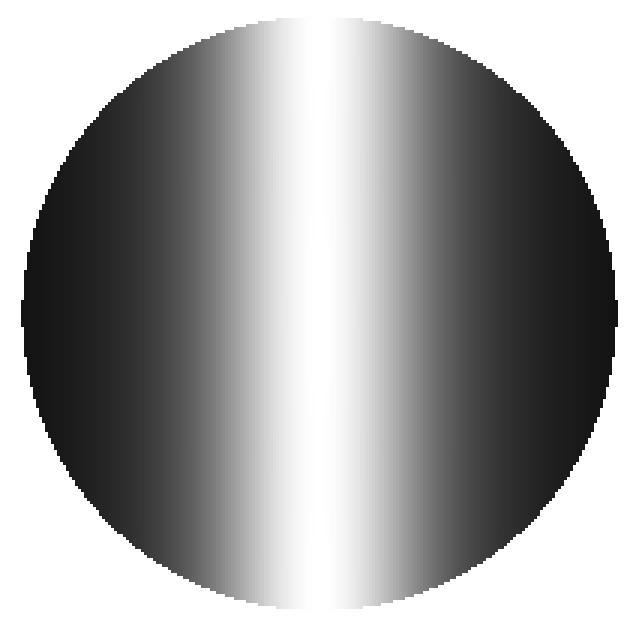}
    \end{minipage}}     
  \subfloat[7\fdg5]{
    \begin{minipage}[c]{\tilewidth}
      \includegraphics[width=\linewidth]{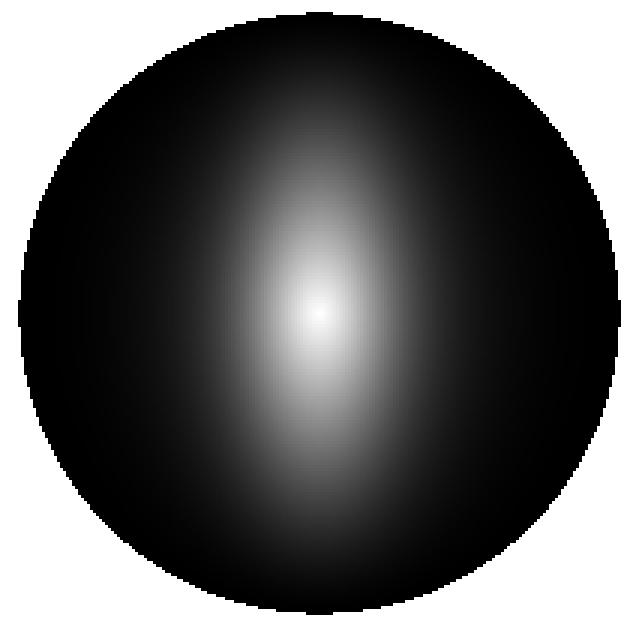}\\
      \includegraphics[width=\linewidth]{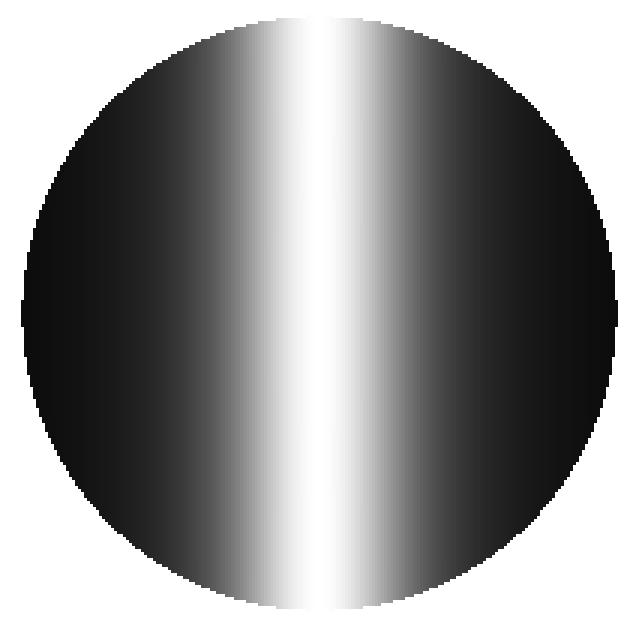}
    \end{minipage}}     
  \caption{\textbf{Top:} Diffraction limited MTFs modified by the
    filter tilt effects, $S_0\cdot\abs{\Stilt(\theta)}$;
    \textbf{Bottom:}~Tilt MTFs, $\abs{\Stilt(\theta)}$. Tilt angles
    $\theta$ as indicated.}
  \label{fig:mtfs}
\end{figure}

The computed MTFs shown in Fig.~\ref{fig:mtfs} demonstrate that the
tilt is a one-dimensional effect; there is no effect at all in the $y$
direction, perpendicular to the tilt.

\begin{figure}[!t]
  \def\tilewidth{0.15\linewidth}
  \subfloat[0\fdg0]{
    \begin{minipage}[c]{\tilewidth}
      \includegraphics[width=\linewidth]{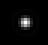}\\
      \includegraphics[width=\linewidth]{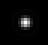}
    \end{minipage}}     
  \subfloat[1\fdg5]{
    \begin{minipage}[c]{\tilewidth}
      \includegraphics[width=\linewidth]{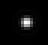}\\
      \includegraphics[width=\linewidth]{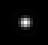}
    \end{minipage}}     
  \subfloat[3\fdg0]{
    \begin{minipage}[c]{\tilewidth}
      \includegraphics[width=\linewidth]{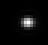}\\
      \includegraphics[width=\linewidth]{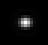}
    \end{minipage}}     
  \subfloat[4\fdg5]{
    \begin{minipage}[c]{\tilewidth}
      \includegraphics[width=\linewidth]{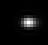}\\
      \includegraphics[width=\linewidth]{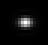}
    \end{minipage}}     
  \subfloat[6\fdg0]{
    \begin{minipage}[c]{\tilewidth}
      \includegraphics[width=\linewidth]{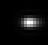}\\
      \includegraphics[width=\linewidth]{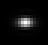}
    \end{minipage}}     
  \subfloat[7\fdg5]{
    \begin{minipage}[c]{\tilewidth}
      \includegraphics[width=\linewidth]{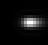}\\
      \includegraphics[width=\linewidth]{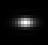}
    \end{minipage}}     
  \caption{\textbf{Top:} PSFs corresponding to
    $S_0\cdot\Stilt(\theta)$; \textbf{Bottom:}~PSFs calculated with
    $\arg\{\tp\}$ set to zero. Tilt angles $\theta$ as indicated.}
  \label{fig:psfs}
\end{figure}

The PSFs corresponding to large angles shift significantly in the $x$
direction, as can be seen in Fig.~\ref{fig:psfs}. This indicates that
there is a significant common tilt in the $x$ direction in the
monochromatic wavefronts. This can be understood by noting that in
Fig.~\ref{fig:barr_transmission_cube}, the monochromatic phases have
gradients in the same positions that there are peaks in the modulus.
Ignoring the phases produces PSFs that are not shifted and also do not
contain the asymmetry in shape that is also evident in the figure.

In real data, we see much larger shifts than the few pixels visible in
Fig.~\ref{fig:psfs}. In fact, most of the shift is caused by
refraction through the glass substrate. An order of magnitude estimate
of this shift can be calculated by use of the formula $d =
t\theta(n-1)/n$ given by \citet[Chapter 4]{smith90modern}. With
$n=1.596$, thickness $t\approx5$~mm, and tilt angle $\theta=6\fdg0=
0.10$~rad, we get a shift of $d \approx 0.20~\text{mm} = 26$~pixels of
size 7.4~$\muup$m.

\begin{figure}[tb]
  \centering
  \includegraphics[bb=54 80 431 254,width=\linewidth]{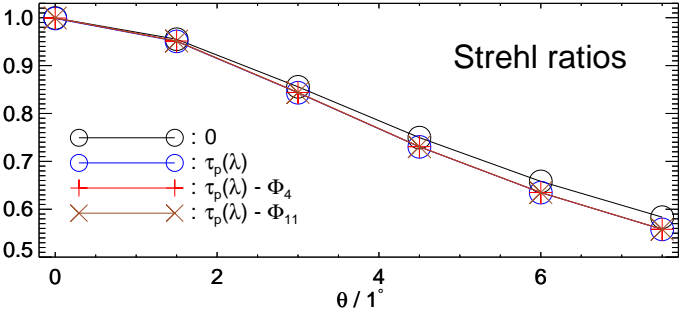}
  \caption{Theoretical Strehl ratios for the \ion{Ca}{ii}~H filter,
    calculated by integrating the MTFs. The pupil functions at each
    wavelength are calculated without the phase, with the phase, and
    with two different compensations for the phase. $\Phi_4$ is an
    optimum focus term, while $\Phi_{11}$ is an optimized sum of
    focus, astigmatism, coma, trifoil, and spherical aberrations.}
  \label{fig:strehl}
\end{figure}

\citet{scharmer06comments} found an optimum focus compensation for the
phase effect of a non-tilted FPI by maximizing the Strehl ratio. In
Fig.~\ref{fig:strehl}, we show Strehl ratios calculated with and
without taking the phase of the transmission into account. The Strehl
ratio without phase is reduced from almost unity for $\theta=0$ to
$\sim$0.6 for 7\fdg5. Including the phase reduces the Strehl further,
particularly for larger $\theta$. Maximizing the Strehl ratio by
varying a common focus compensation we find negligible improvements in
Strehl ratio. A tenth of a percent for the small angles, even less for
the larger angles. The wavefront RMS of the estimated focus terms
ranges from 0.010 waves for $\theta=0\fdg0$ to 0.006 waves for
$\theta=7\fdg5$.

The individual apodisations and phases for the non-zero angles are not
circularly symmetrical so astigmatism and coma would not be
unexpected. However, including Zernike polynomials 4--11 in the
optimization gives insignificant additional Strehl improvements and no
change in the estimated wavefront RMS.\footnote{We used Zernike
  polynomials ordered as specified by \citet{noll76zernike}, where
  indices 4--11 correspond to focus, astigmatism, coma, trifoil, and
spherical aberrations. For the 1D optimization of the focus term
alone, we used Brent's Method. Starting from the focus-only optimum,
we used the Downhill Simplex Method (AMOEBA) for Zernikes 4--11. For
both methods, see \cite{press86numerical}.} Changes, if any, are in the
sixth digit. There is no reason to expect higher order modes to have a
significant effect.

The compensated Strehl ratios are also plotted in
Fig.~\ref{fig:strehl}, where $\Phi_M=\sum_{m=4}^M c_m Z_m$ are the
optimized phases. $Z_m$ are Zernike polynomials, $m=4$ corresponds to
focus. It is apparent that the compensations do not alter the Strehl
ratios significantly.

This suggests that, at least for this particular setup, the effect of
the varying incident angles on the filter should interfere very little
with our normal Multi-Object Multi-Frame Blind Deconvolution
\citep[MOMFBD; ][]{noort05solar} restoration practice. Image shifts
are already routinely taken care of in calibration of camera alignment
by use of a pinhole array. The focus term is compensated already when
the camera is focused, usually for $\theta=0\fdg0$. This leaves an
uncompensated focus of only 0.004 waves or less at the larger angles,
which is insignificant compared to the accuracy of our camera
focusing. Evident from the way the PSF moves with angle when the phase
is included is that there is also a global tilt term that varies with
$\theta$, but this does not change the shape of the PSF.

This means the OTF is to a good approximation separable into an
atmospheric wavefront part and a correction for the filter tilt as in
Eqs.~(\ref{eq:18})--(\ref{eq:19}). The smearing caused by the
elongated PSFs affects the SNR by lowering the MTF but should
otherwise pass through the MOMFBD processing without changing the
estimated phases. The restored object is a version of the real object
that is convolved with the PSF corresponding to the apodisation
effects without the wavefront phase. This can be taken care of with a
simple post-restoration deconvolution step. The consequences for
alignment are discussed in Sect.~\ref{sec:shift} below.

\section{Synthetic solar data and expected errors}
\label{sec:synthetic-data}

The Strehl experiments correspond to objects that are point-like and
do not depend on $\lambda$. However, the solar structures vary with
wavelength within the passband of the filter, particularly if it's
broadened and particularly where the line gradient is large. In this
section we simulate the formation of solar images by summing the
contributions from many wavelengths in a synthetic data cube. We make
``perfect'' images as well as images based on convolution with either
$S_0$ or $\Smono$ before summing. We evaluate how well the ``perfect''
data can be recreated by deconvolving the summed images.

\begin{figure*}[!t]
  \centering
  \def\figdir{/home/mats/data/pupil-apodisation/artificial/}
  \subfloat[\label{fig:RMS}]{\includegraphics[bb=39 63 432 272,width=0.49\linewidth]{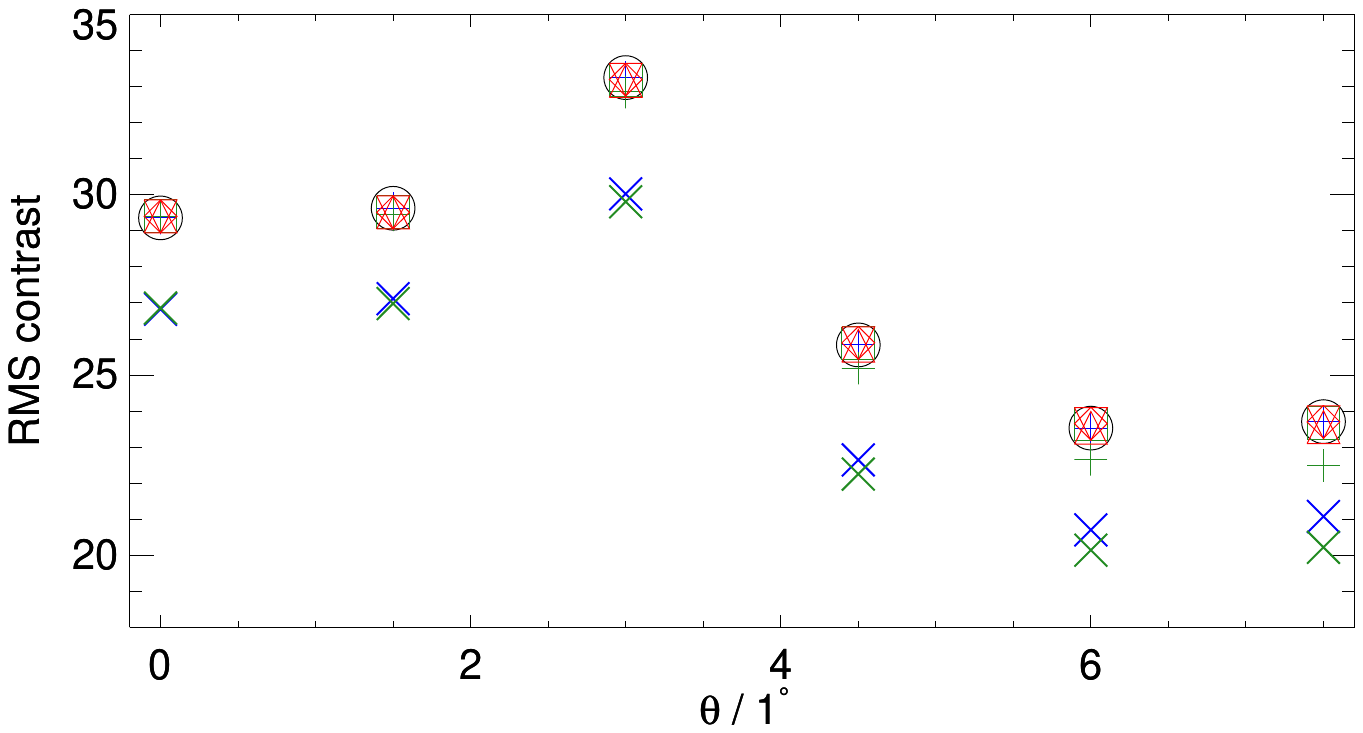}}
  \hfill
  \subfloat[\label{fig:RMSE}]{\includegraphics[bb=39 63 432 272,width=0.49\linewidth]{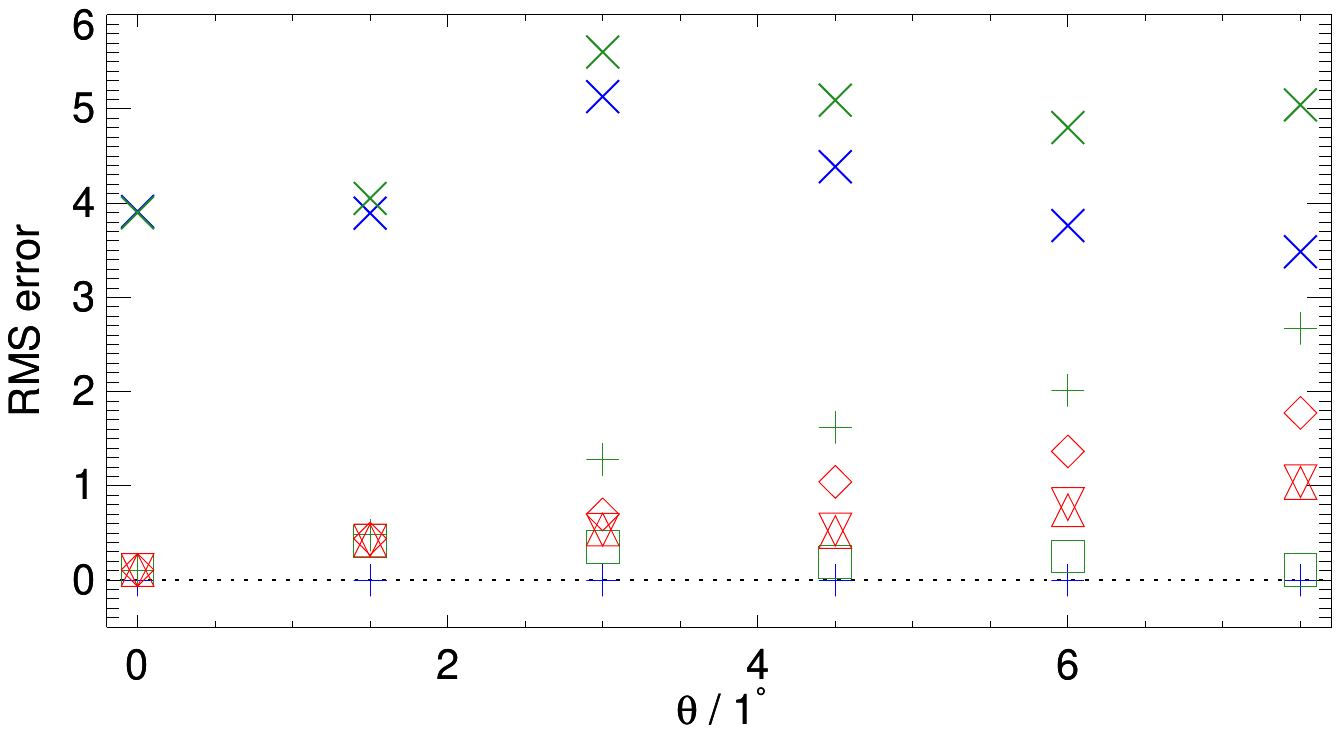}}
  \caption{RMS contrasts (a) and RMS errors (b) from synthetic image
    deconvolution experiment, both in percent of the mean intensity.
    Black circles ($\medcirc$, in a only): Synthetic ``true'' images.
    Blue symbols: monochromatic images convolved with $S_0$ and then
    summed. The crosses ($\times$) correspond to those images and the
    pluses ($+$) to the same images after deconvolution with $S_0$.
    Green symbols: monochromatic images convolved with
    $\Smono(\lambda,\theta)$. The crosses ($\times$) correspond to
    those images and the pluses ($+$) to the same images after
    deconvolution with $S_0$. The squares ($\square$) correspond to
    those summed images again, but now deconvolved with
    $S_0\cdot\Stilt(\theta)$. Red symbols: convolved images like for
    the green symbols but with three types of imperfections in the
    deconvolution. Up triangles ($\bigtriangleup$): transmission phase
    not used for calculating \Stilt. Down triangles
    ($\bigtriangledown$): deconvolve with the MTF,
    $S_0\cdot\abs{\Stilt(\theta)}$. Diamonds ($\Diamond$):
    deconvolution with $S_0\cdot\Stilt(\theta)$ mirrored in the $x$
    direction.}
  \label{fig:RMSandRMSE}
\end{figure*}

The synthetic image data cube was calculated from 3D MHD simulations
kindly provided by Mats Carlsson and based on the code by
\citet{1998ApJ...499..914S}. We used MULTI
\citep{1986UppOR..33.....C}, a \ion{Ca}{ii}~H 5-level atom model, and
line opacities from the Vienna Atomic Line Database
\citep{1995A&AS..112..525P,1999A&AS..138..119K,2000BaltA...9..590K}
for the line blends. The data cube comprises a wavelength range
covering the tunable filter range ($394.2\text{
  nm}\le\lambda\le399.5\text{ nm}$, $\delta\lambda=0.01\text{ nm}$).
The most relevant part of the spectrum is shown in
Fig.~\ref{fig:profiles}.

Because we have established separability in
Sect.~\ref{sec:point-spre-funct}, we do not have to include
atmospheric turbulence in these simulations. The experiment
corresponds to images that are perfectly restored from those effects
by image restoration with methods such as MOMFBD or Speckle
interferometry. We included neither noise nor stray light in these
simulations.

The results of the experiments can be found in
Fig.~\ref{fig:RMSandRMSE}. The RMS errors are calculated as the
standard deviation of the difference of an image and the ``perfect''
image, after subpixel alignment. The errors therefore measure only the
residual blurring and not errors from image shifts caused by
off-center and asymmetrical PSFs.

The circles in Fig.~\ref{fig:RMS} represent the RMS contrasts of the
``true'' images. These contrasts vary between 23.5\% and 33.3\%
because of the varying heights in the solar atmosphere where they are
formed. Convolving each monochromatic image in the synthetic data cube
with the diffraction-limited PSF before summing, i.e., multiplying the
Fourier transform with $S_0$, makes the contrasts drop as shown with
the blue crosses, resulting in RMS errors in the range 4--5\% as shown
in Fig.~\ref{fig:RMSE}. Deconvolving with $S_0$ recreates the original
contrasts and brings the errors down to zero as expected (blue
pluses).

For the remaining experiments, $\Smono$ was applied to each
monochromatic image before summing. The contrast drops a bit more than
for $S_0$, particularly for the larger tilt angles $\theta$ and the
RMS errors are larger (green crosses). Consequently, deconvolving with
$S_0$ works well for the small angles but not for the larger angles
(green pluses). In fact, for 0\fdg0 and 1\fdg5, it does not matter
much whether we deconvolve with $S_0$, $\Stilt$ or any of the
imperfect OTFs described in the next paragraph.

For the larger angles, deconvolving with $S_0\cdot\Stilt$ also
restores the original contrasts and reduces the errors to small
fractions of the $\Smono$ versions. However, when using imperfect
versions of $\Stilt$ (red symbols) the results for the larger angles
is worse. Not using the phase of the transmittance profile,
$\arg(\tf)$, gives approximately the same RMS errors as correcting
only for the MTF, $S_0\cdot\abs{\Stilt}$, which is expected since
neither can represent the asymmetry of the PSFs (red upward and
downward triangles, resp.). Even worse is to use the asymmetric PSFs,
but reversed (red diamonds).

\section{Real data and consequences for image restoration}
\label{sec:real-data}

In this section, we work with SST data collected by Henriques et
al. (in prep.)  in May 2010. We used four MegaPlus~II es4020 cameras
with 2048$\times$2048 7.4-$\muup$m pixels. The beam was F/46 and the
image scale is 0\farcs034/pixel. There was 23\% oversampling in the
Fourier domain.

One camera collected data throught the tilt-mounted NB filter. The
observing programme specified tilt angles $\theta=1\fdg2$, 2\fdg5,
3\fdg0, 3\fdg7, 4\fdg2, 4\fdg8, and 6\fdg4. These angles correspond to
passbands with the central wavelengths $\lambda_\text{c}=396.84$,
396.74, 396.67, 396.57, 396.47, 396.34, and 395.93~nm, respectively,
chosen to sample interesting heights in the solar atmosphere. In
addition, we had two cameras behind a wideband (WB) filter centered on
395.37~nm (FWHM 1.0~nm), one in the conventional focus and one
defocused by 7 mm for approximately 1~wave peak-to-peak of focus PD.
The fourth camera was mounted behind a fixed NB (FWHM 0.1~nm)
396.47~nm wing filter.

We routinely perform a calibration step, where we collect data while
scanning the filter through a large range of angles, approximately
centered on $\theta=0\degr$ (in the case of these data: 20\degr{} with
0\fdg1 resolution). We calculate the average intensities of the images
and find the zero position by looking for the symmetry point. Because
these data depend on the unknown spectrum\footnote{The Liege and FTS
  atlases differ and neither is guaranteed to match the data at our
  position.}, we do not model these data in order to draw any
conclusions besides the zero tilt angle.

\subsection{Pinhole images and OTF correction}
\label{sec:pinholes}

As part of the standard calibration procedures at the SST, we collect
data through artificial targets mounted in the primary (Schupmann)
focus. One such target, designed for measuring straylight, consists of
six holes of diameters that range from 20~$\muup$m to 1~mm. We can use
these to test our calculated \Stilt{} corrections. We use a
512$\times$512-pixel subfield with only the smallest, barely resolved
pinhole.

In order to gain SNR, we co-added many frames (each with an exposure
time of 10~ms), aligned to subpixel precision. The available number of
frames from this campaign is 500 in each of the tilted NB wavelengths
and 3500 in the WB. The individual frames are dark corrected by
subtraction of an average dark frame. After co-adding, an extra dark
correction step is performed. All summed images were normalized to the
average intensity within the 1-mm hole. A refinement dark level for
the WB image was calculated as the median of the outer few rows and
columns of the 512$\times$512-pixel subfield and subtracted from the
WB image. The NB images were then dark corrected by subtraction of the
amount needed to make the total intensity the same in all images.

\begin{figure}[!t]
  \centering
  \def\tilewidth{\linewidth}
  \includegraphics[bb=50 11 491 342,width=\tilewidth]{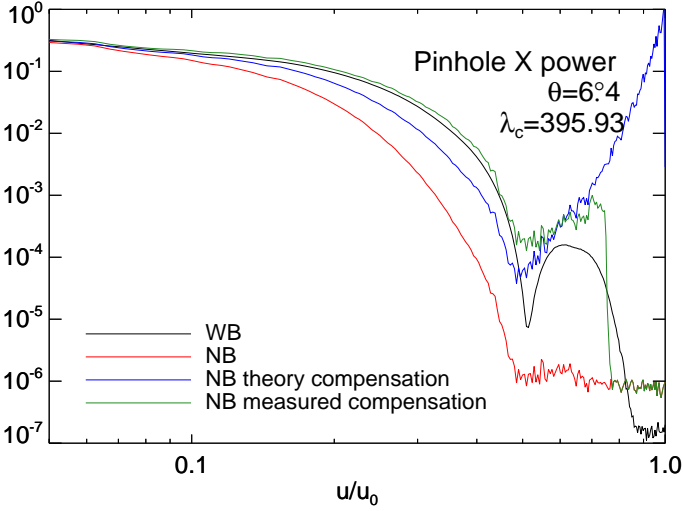}
  \caption{Straylight target pinhole power spectra in the $x$
    direction, affected by the 6\fdg4 tilt. Angular averages within
    $\pm22.5\degr{}$ wide sectors centered on the axis directions.
    Calculated using a 1024$\times$1024-pixel subfield of the detector
    FOV. NB: original, corrected with theoretical MTF, and corrected
    with measured MTF (or OTF, the corrected powers are the same). WB
    for comparison (covered by the NB compensated line). The TRIPPEL
    measured profile was used for the theoretical correction. Using
    the theoretical profile gives even less correction.}
  \label{fig:pinhole-power}
\end{figure}

The power of the pinhole image is isotropic in the WB but not in the
NB because of the tilt effect. In the $y$ direction the WB and NB
powers are equal but in the $x$ direction, the NB power is attenuated.
We demonstrate this for $\theta=6\fdg4$ in
Fig.~\ref{fig:pinhole-power}, compare the black and red curves. Note
the dip in power at approximately 50\% of the diffraction limit. There
is a corresponding dip in the NB power, although not as clearly
visible because the power spectrum is so noisy at the spatial
frequencies beyond the dip. These dips are caused by a zero-crossing
in the Fourier transform of the target. The Fourier transform of a
pillbox function (representing a circular pinhole) is a radial
jinc\footnote{Similar to a sinc function, $\jinc(r)=J_1(r)/r$, where
  $J_1$ is a Bessel function of the first kind, oscillates with
  decreasing amplitude but without a defined period.} function. Such
zero crossings appear closer to the origin the larger the hole is, so
the smaller the hole, the better. In the limit of zero diameter we
have a $\delta$ function, the transform of which is a constant.

Correcting the NB power by use of the theoretical $\Stilt$ is not
enough, as evidenced by the blue curve. We show here the theoretical
correction based on the profile measured with TRIPPEL. Using the
manufacturer's theoretical profile results in even slightly less
correction.

\begin{figure}[t]
  \centering
  \def\tilewidth{0.23\linewidth}
  \def\figdir{/home/mats/data/pupil-apodisation/stray}
  \subfloat[6\fdg4]{
    \begin{minipage}[c]{\tilewidth}
      \includegraphics[angle=90,width=\linewidth]{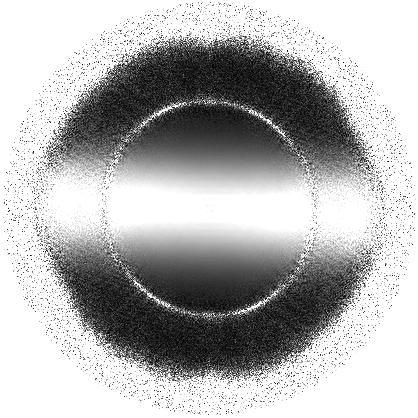}
      \includegraphics[angle=90,width=\linewidth]{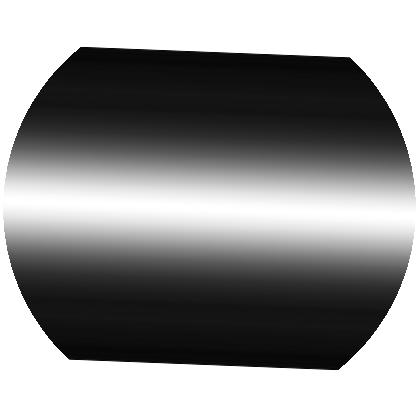}
    \end{minipage}}
  \subfloat[3\fdg0]{
    \begin{minipage}[c]{\tilewidth}
      \includegraphics[angle=90,width=\linewidth]{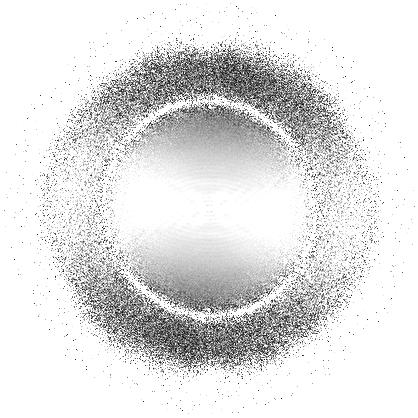}
      \includegraphics[angle=90,width=\linewidth]{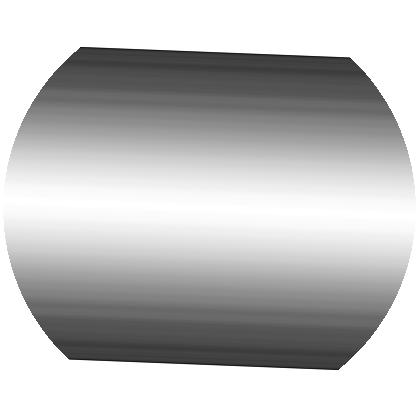}
    \end{minipage}}
  \subfloat[2\fdg5]{
    \begin{minipage}[c]{\tilewidth}
      \includegraphics[angle=90,width=\linewidth]{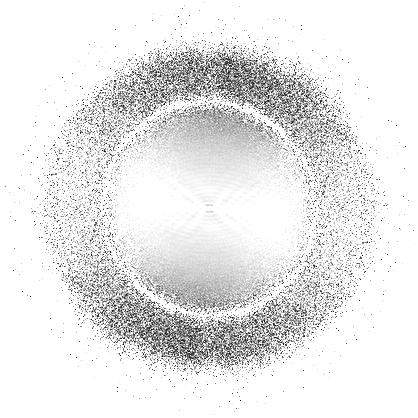}
      \includegraphics[angle=90,width=\linewidth]{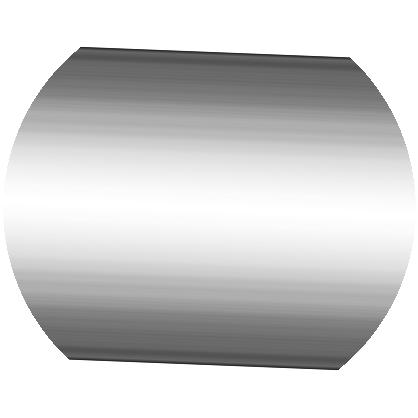}
    \end{minipage}}
  \subfloat[1\fdg2]{
    \begin{minipage}[c]{\tilewidth}
      \includegraphics[angle=90,width=\linewidth]{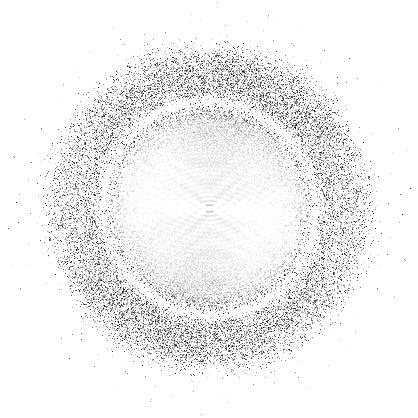}
      \includegraphics[angle=90,width=\linewidth]{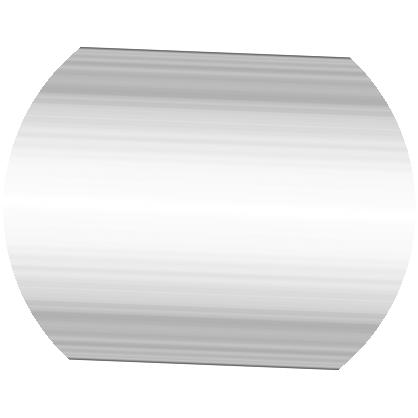}
    \end{minipage}}
  \caption{Measured MTF correction $\abs{\Stilt(\theta)}$ based on
    straylight target pinhole power. Top: Dirty; Bottom: Cleaned. Tilt
    angles $\theta$ as indicated. The circumferences represent the
    diffraction limit. The orientation of the ridge is $\sim$2\degr{}
    from the $y$ axis. Compare with Fig.~\ref{fig:mtfs}.}
  \label{fig:MTFcorrection}
\end{figure}

However, because the pinhole is the same in WB and NB, it should be
possible to measure \Stilt. We can get a first estimate by taking the
ratio of the Fourier transforms of the two images,
\begin{equation}
  \label{eq:10}
  \frac{D_\text{NB}}{D_\text{WB}}
  = \frac{F \Swf \Stilt}{F\Swf} 
  = S_\text{tilt},
\end{equation}
where $\Swf$ is here the OTF representing phase aberrations in the
optics on the optical table. The results for a few tilt angles are
shown in the top row of Fig.~\ref{fig:MTFcorrection}. These ``dirty''
measurements are dominated by noise at the higher spatial frequencies.
However, since we know from Fig.~\ref{fig:mtfs} that they should be
constant in the $y$ direction, it should be possible to clean them.

The ring at about 50\% of the diffraction limit corresponds to the
jinc dip in WB power. We mask this ring-shaped zero crossing artifact
and calculate the median along the direction that is supposed to be
constant. We smooth the result and then construct a cleaned 2D
\Stilt{}. In the real data, this direction is not exactly parallel to
the $y$ direction, so we first find the orientation of the ridge,
rotate \Stilt{}, do the cleaning and rotate the result back to the
original orientation. The result is in the bottom row of
Fig.~\ref{fig:MTFcorrection}.

\begin{figure}[!t]
  \includegraphics[bb=75 34 465 218,clip,width=\linewidth]{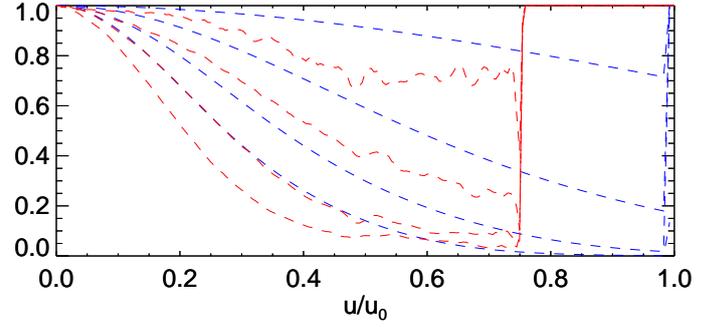}
  \caption{MTF corrections, $\abs{\Stilt(\theta)}$, in the $x$
    direction for $\theta=1\fdg2, 3\fdg0, 4\fdg8, 6\fdg4$ (from top to
    bottom). Blue: theoretical MTFs; Red: measured MTFs.}
  \label{fig:MTF-plots-barr}
\end{figure}

We show the theoretical and measured corrections for a few angles in
Fig.~\ref{fig:MTF-plots-barr}. Note the discrepancies between the blue
curves and the corresponding red curves! For the present data, the
limit to where we can measure \Stilt{} in the $x$ direction is set by
the SNR in the WB pinhole image in the denominator to a little less
than 80\% of the diffraction limit. We define \Stilt{} to be unity
outside this limit. Noise in the NB (numerator) is worse for the
smaller angles because the Sun is darker in the core of the spectral
line, the irregularities caused by this are apparent in the figure.

It is apparent that we need more exposures to construct \Stilt{}
functions that go all the way to the diffraction limit in the $x$
direction, and that are not noise dominated above 50\% of the
diffraction limit. The noisy NB bump in Fig.~\ref{fig:pinhole-power}
suggests that we need to increase the number of NB frames by an order
of magnitude or more. With an image rate of 10~frames/s, it should be
possible to collect $\sim$5000 images in 8~min. With seven tilt
angles, this corresponds to almost an hour. However, this kind of data
would not have to be collected each day or even by each observer
interested in this kind of correction.

\subsection{Solar images}

\begin{figure*}[!t]
  \centering
  \def\figdir{/home/mats/data/pupil-apodisation/solar/data_subfielded}
  \def\tilewidth{0.45\linewidth}
  \subfloat[\label{fig:solpowx}]{\includegraphics[bb=50 11 491 342,width=\tilewidth]{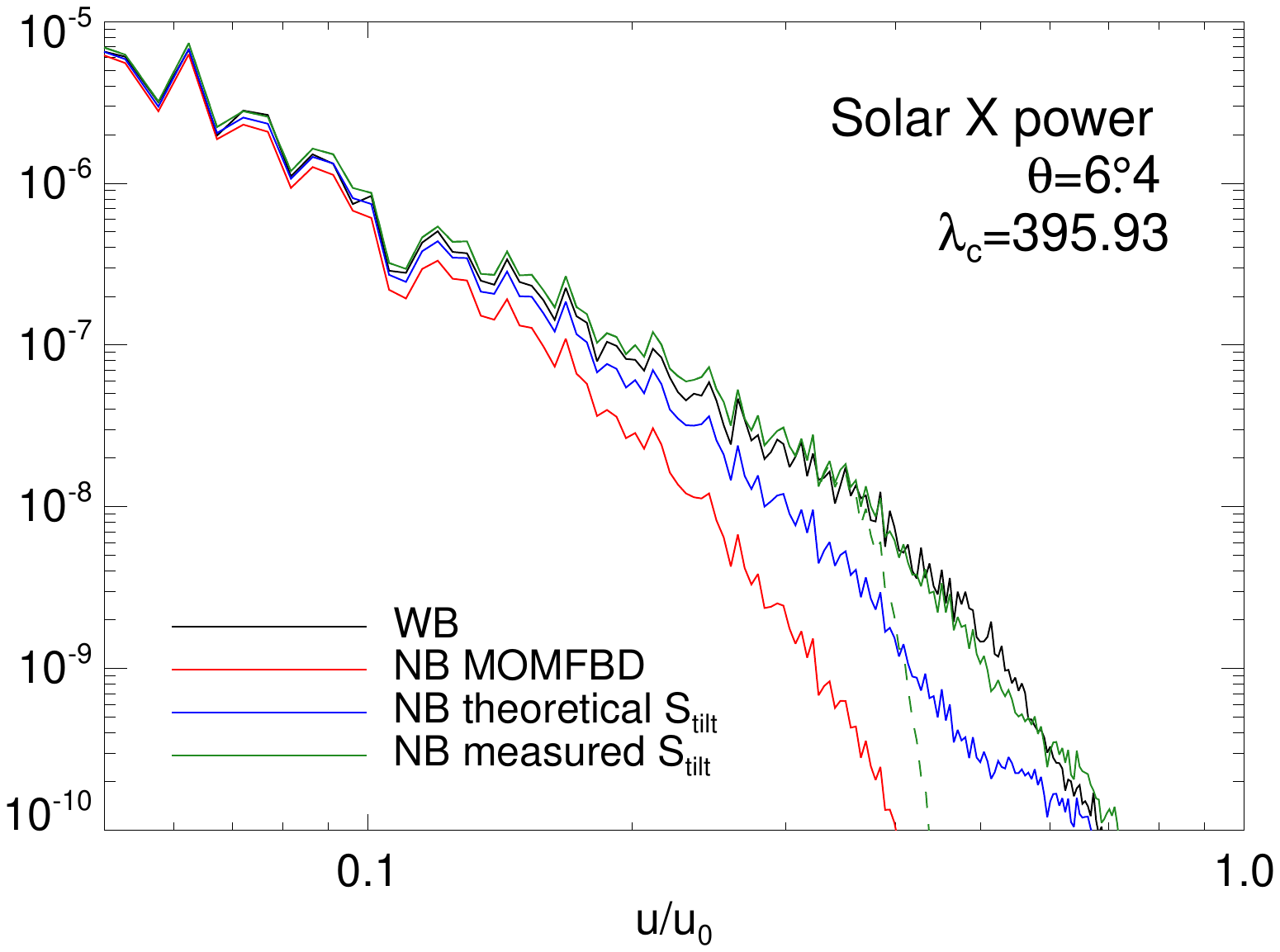}}\quad
  \subfloat[\label{fig:solpowy}]{\includegraphics[bb=50 11 491 342,width=\tilewidth]{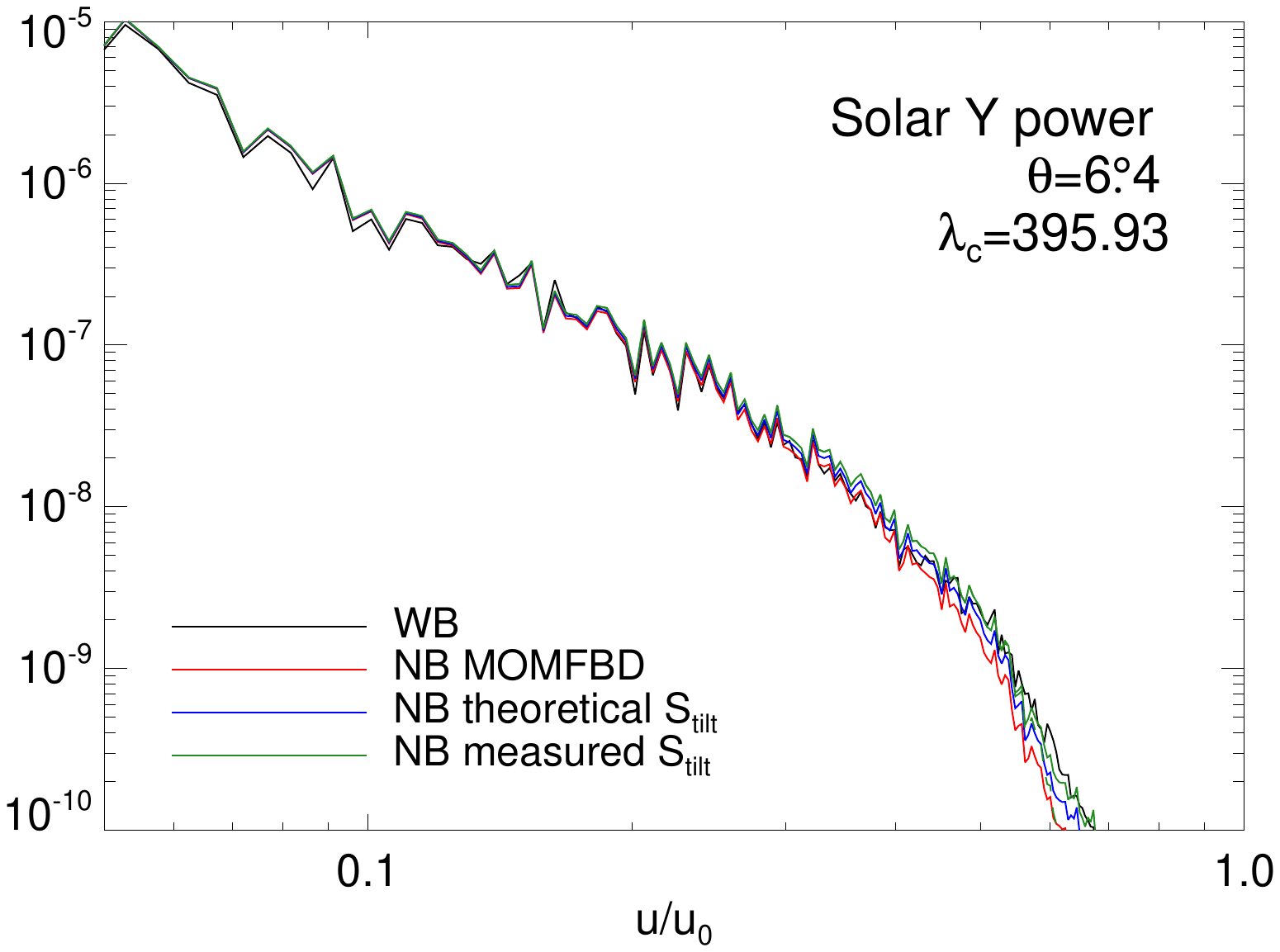}}
  \caption{Solar power spectra, angular averages within
    $\pm22.5\degr{}$ wide sectors centered on the axis directions.
    \textbf{\protect\subref{fig:solpowx}} In the $x$ direction, where the tilt
    effects are; \textbf{\protect\subref{fig:solpowy}} In the $y$ direction,
    unaffected by the tilt effects. The dashed line represents noise
    filtered data.}
  \label{fig:solar-power}
\end{figure*}

On 2010-05-23, we collected solar images with an exposure time of
10~ms near disc center ($\mu\approx\cos 20\degr\approx0.94$). Together
with the WB, WB PD, and fixed NB data, the tilt-tuned NB images were
restored for atmospheric turbulence effects with MOMFBD using the 36
most significant atmospheric Karhunen--Lo\`eve modes. Observations and
MOMFBD processing will be described in detail by Henriques et al. (in
prep.).

Figure~\ref{fig:solar-power} shows power spectra for $\theta=6\fdg4$.
For this angle, the central wavelength of the NB passband has shifted
to 395.93~nm, well within the passband of the WB filter. Here, the
granulation has the same power in the $y$ direction in NB as in WB,
see Fig.~\ref{fig:solpowy}. We can expect this to be the case also in
the $x$ direction, which can be used for testing our \Stilt{}
compensation. In Fig.~\ref{fig:solpowx}, the MOMFBD restored NB power
(red) is attenuated compared to the WB power (black). Just as for the
pinholes, the theoretical \Stilt{} compensation (blue) does not fully
correct the asymmetry in power, but the measured compensation does
(green).
 
\begin{figure*}[!t]
  \centering
  \def\tilewidth{0.33\textwidth}
  \subfloat[\label{fig:solar-images-momfbd}]{\includegraphics[width=\tilewidth]{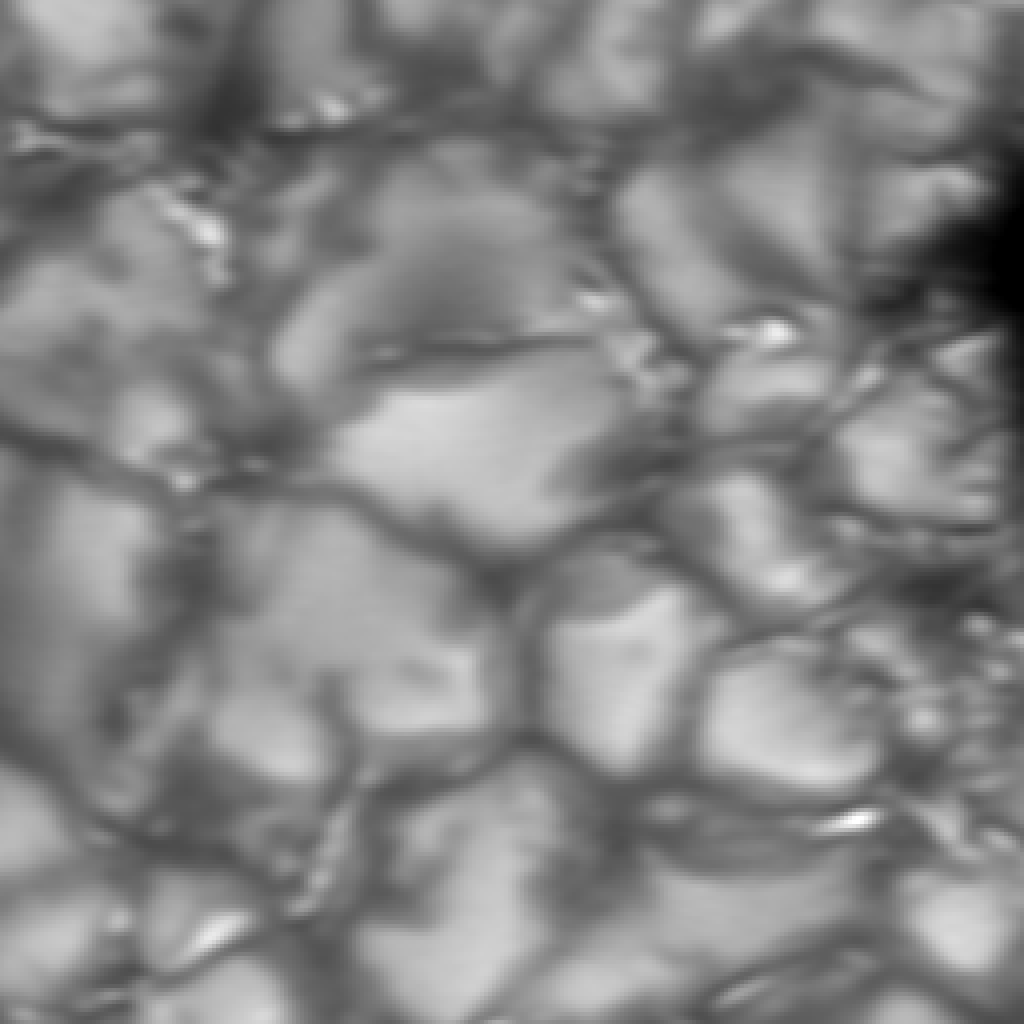}}\hfill
  \subfloat[\label{fig:solar-images-comp}]{\includegraphics[width=\tilewidth]{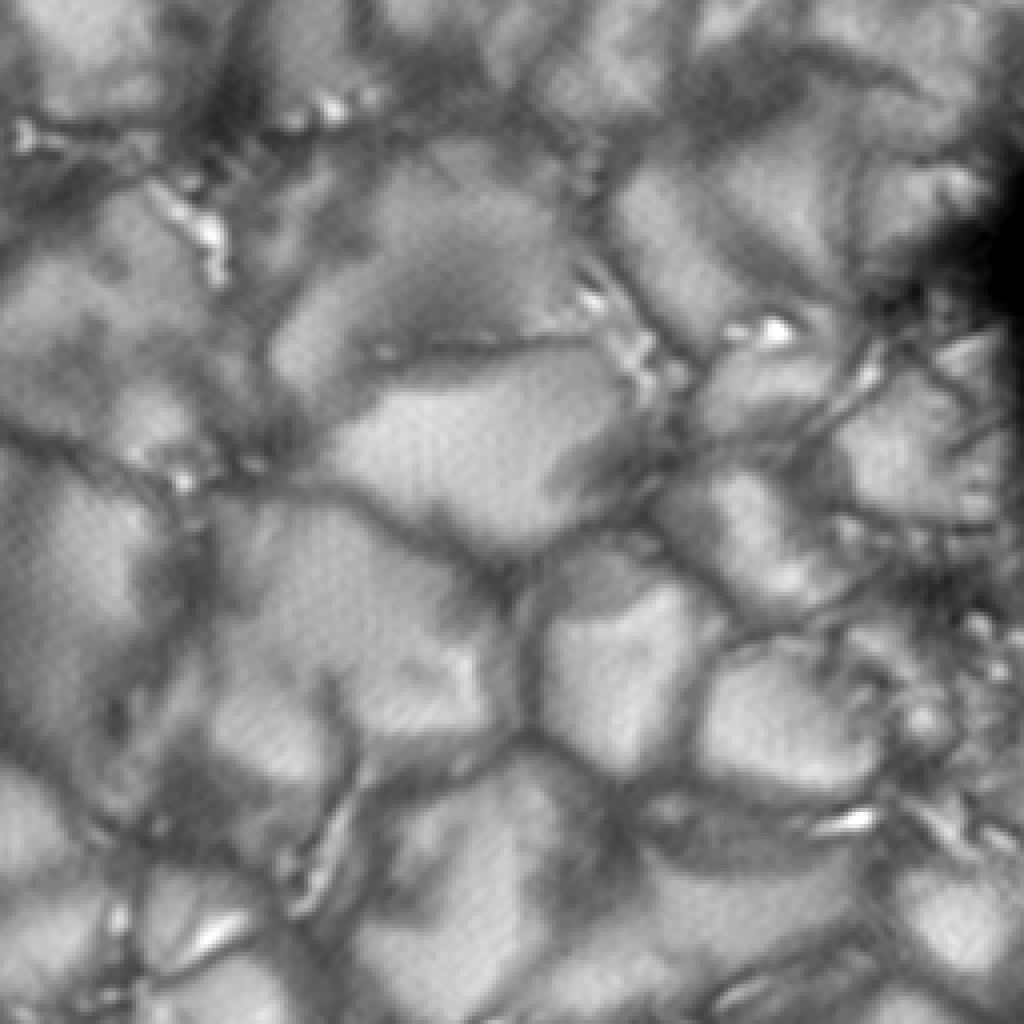}}\hfill
  \subfloat[\label{fig:solar-images-filt}]{\includegraphics[width=\tilewidth]{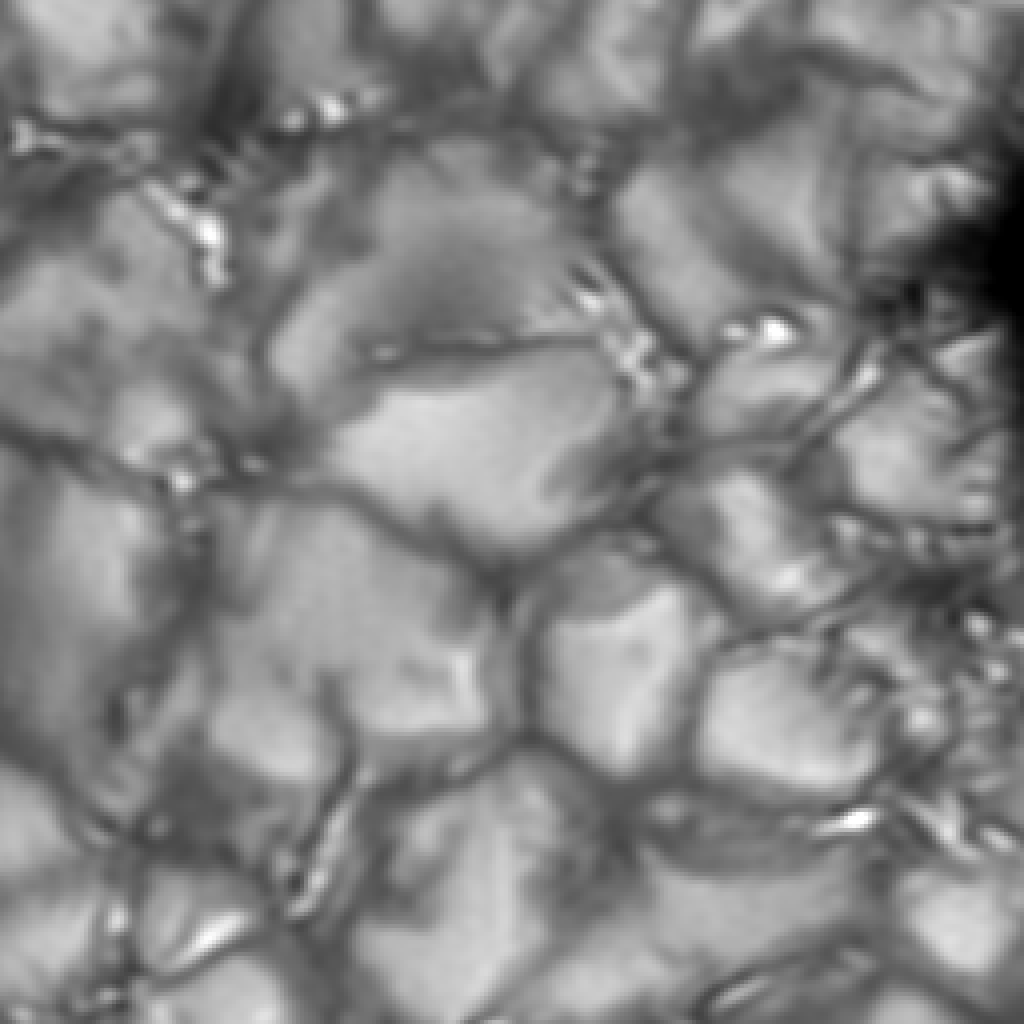}}\\
  \caption{An 11\arcsec$\times$11\arcsec{} solar image collected with
    6\fdg4 filter tilt. \textbf{(a)}~Restored with MOMFBD;
    \textbf{(b)}~Compensated for tilt effects by use of the measured
    \Stilt; \textbf{(c)}~Compensated and noise filtered.}
  \label{fig:solar-images}
\end{figure*}

There are visible effects in the images, that correspond to the
correction of the anisotropy in the power spectrum. Compare the MOMFBD
restored image in Fig.~\ref{fig:solar-images-momfbd} and the same
image after correction for \Stilt{} in
Fig.~\ref{fig:solar-images-comp}.  Because \Stilt{} makes point-like
objects elongated in the $x$ direction, the effect is most easily seen
in the smallest structures. For example, strings of small bright
features oriented in the $x$ direction are more clearly separated
after \Stilt{} correction, while such strings oriented in the
perpendicular direction mainly get higher contrast with respect to the
surrounding dark lanes.

\begin{figure}[!t]
  \centering
  \def\tilewidth{0.3\linewidth}
  \subfloat[6\fdg4\label{fig:noise-filter64}]{\includegraphics[width=\tilewidth]{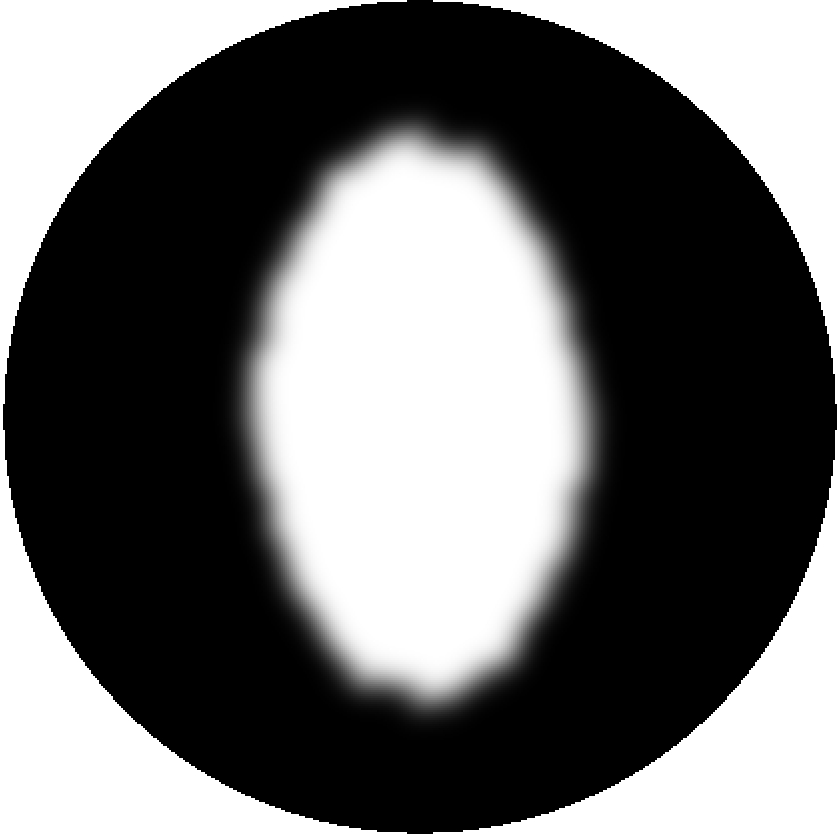}}\
  \subfloat[4\fdg8\label{fig:noise-filter48}]{\includegraphics[width=\tilewidth]{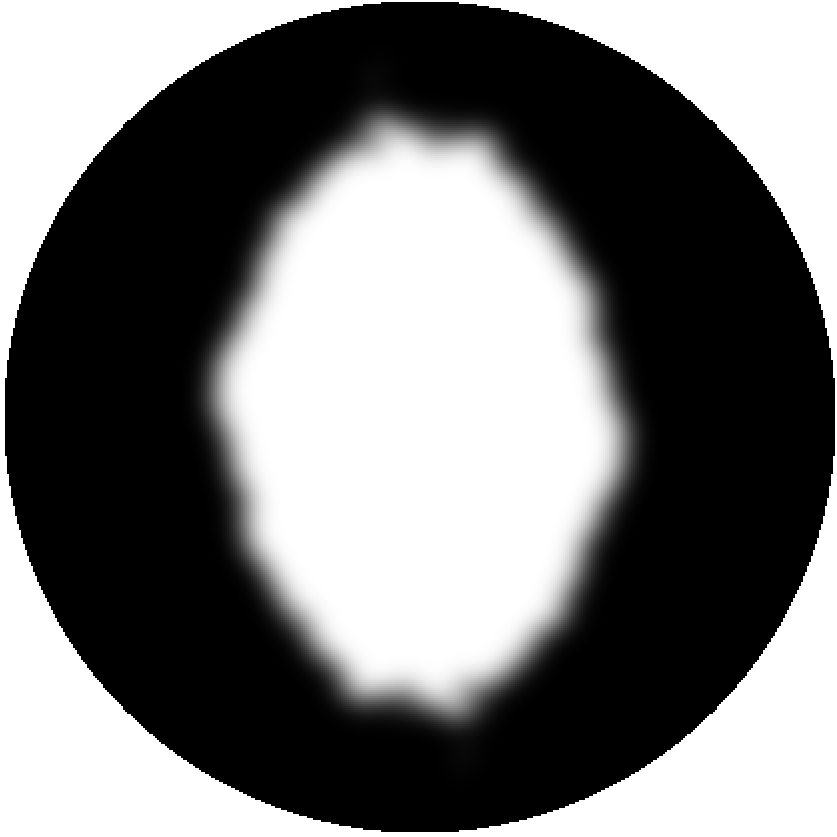}}\
  \subfloat[4\fdg2\label{fig:noise-filter42}]{\includegraphics[width=\tilewidth]{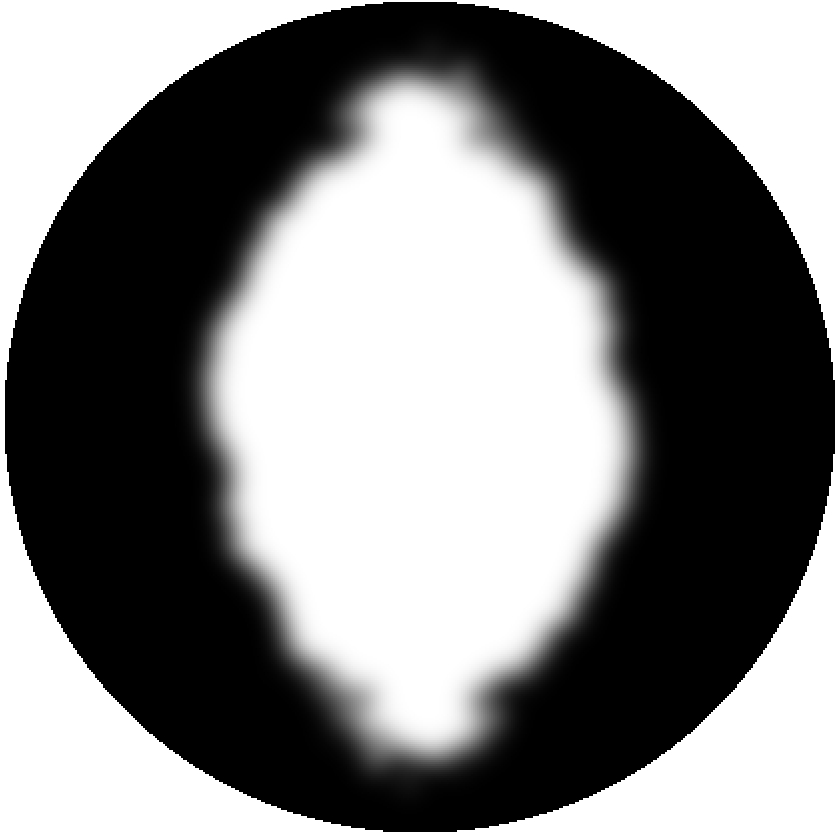}}
  \caption{The low-pass noise filter for the largest angles. The 
    perimeter of the circle represents the diffraction limit.}
  \label{fig:noise-filter}
\end{figure}

The \Stilt{} compensation enhances the noise somewhat, particularly
for the larger angles, so we need to use a noise filter.
Figure~\ref{fig:solar-images-filt} shows the compensated image
filtered with the noise filter in Fig.~\ref{fig:noise-filter64}. The
noise filter is very asymmetrical, as it should because the attenuated
power in the $x$ direction will fall to the noise level faster than
the $y$ power. The noise filter seems to be correct in that it removes
the noise without visibly changing the resolution in the smallest
structures. This is also the case for the noise filters constructed
for the other tilt angles.
\begin{table*}[!t]
  \centering
  \caption{Observed RMS contrasts.}
  \footnotesize  
  \begin{tabular}{lrrrrrrrr}
    \hline
    \hline\noalign{\smallskip}
    \multirow{2}{*}{Deconvolution} & \multicolumn{7}{c}{NB}&\multirow{2}{*}{WB} \\
    \cline{2-8}\noalign{\smallskip}
    & 1\fdg2 & 2\fdg5 & 3\fdg0 & 3\fdg7 & 4\fdg2 & 4\fdg8 & 6\fdg4\\
    \hline\noalign{\smallskip}
    $\Satm$                &25.7 & 16.5 & 15.4 & 13.1 & 12.0 & 12.2 & 12.9 & 13.3\\
    $\Satm\Stilt(\theta)$  &25.8 & 16.6 & 15.6 & 13.6 & 12.4 & 12.7 & 13.6 & \\
    \hline
  \end{tabular}
  \tablefoot{RMS contrasts of real NB data in percent of the mean
    intensity. The first row 
    has contrasts for images corrected for atmospheric turbulence by
    the MOMFBD program. The second row contrasts are for the same
    images after compensation also for \Stilt. WB contrasts are given
    for comparison with the 6\fdg4 data.} 
  \label{tab:obs_contrasts}
\end{table*}

For a proper Wiener-type noise filter, one needs the signal power and
the noise power. The original, raw-data noise at the highest spatial
frequencies is already filtered out by the MOMFBD
restoration\footnote{In the raw data, assuming additive Gaussian
  noise, the MOMFBD program measures the noise level outside the
  diffraction limit where we know there is no signal.}. Instead, there
is a noise that comes from the mosaicking of subfields done as part of
the MOMFBD processing. The power of this noise seems to vary with
spatial frequency in a way similar to the signal but extend to higher
spatial frequencies. The filters in Fig.~\ref{fig:noise-filter} are
based on thresholding the power spectrum of the uncompensated NB image
at levels found by trial and error. The filters are then made from the
resulting binary images by closing holes and removing isolated pixels,
and finally smoothing with a Gaussian kernel.

For images to be used for making differential quantities (like
magnetograms or Doppler maps), it is best to use the same low-pass
noise filter. Otherwise, there will be spatial frequencies with
information missing in some images but not in others. A simple way of
constructing such a filter is to form, in every Fourier domain pixel,
the minimum of the filters for the relevant images. Such a combination
of the filters shown in Fig.~\ref{fig:noise-filter} would be very
similar to the 6\fdg4 filter.

The contrasts in Table~\ref{tab:obs_contrasts} are much lower than
those in Fig.~\ref{fig:RMSandRMSE} because of various sources of stray
light, see discussion by \citet{scharmer10high-order}. The contrasts
show the expected increase in contrast with $\Stilt$ compensation,
more for large $\theta$ than for small.

\subsection{Image shift and MOMFBD processing}
\label{sec:shift}

A very useful property of the Multi-Object part of MOMFBD image
restoration is that the restored images of the different objects are
delivered by the program, aligned to subpixel precision
\citep{noort05solar}. In this context, different ``objects'' refer to
the same patch of the Sun, but imaged in different wavelengths and/or
different polarization states. This ability facilitates the
calculation of physical quantities based on combinations of these
objects with a minimum of artifacts from misalignment.
When the images are co-spatial and co-temporal, they are subject to
the same wavefront aberrations and are therefore blurred and shifted
by the same PSFs (or PSFs that differ in easily modeled ways). The
alignment is accomplished by a pre-processing step, where local shifts
are measured in images of an array of pinholes, which provides enough
information for the MOMFBD program to compensate for fixed differences
in alignment and, to some extent, in rotation or image scale.

Compensation for \Stilt{} interferes with this procedure because the
corresponding PSF is off-center and therefore shifts the images. If
the MOMFBD-restored images are shifted in a post-processing step, it
is necessary to make sure the pinhole array images are shifted the
same way before the local-shift measurements are done. So the pinhole
array images should be \Stilt{} compensated before the pre-processing
step.

However, the difference in shifts for large and small tilt angles can
be several pixels, see Fig.~\ref{fig:psfs}. Applying the compensation
to the pinhole images means the subfield grid used for the
corresponding objects will be off by the same number of pixels. This
is then corrected for in the post-processing step, where the restored
images are \Stilt{} compensated, but it is likely that the alignment
works better if the subfield grids are aligned during the MOMFBD
process. This can be accomplished if the \Stilt{} PSFs are centered
before they are used. We believe it is enough and possibly safest to
do this to nearest pixel and avoid subpixel operations for this step.

We therefore recommend the following steps, which we have tested with
our data set:
\begin{enumerate}
\item Calculate $\Stilt(\theta)$ for all $\theta$ and center the
  corresponding PSFs to nearest pixel precision.
\item Deconvolve the pinhole array images with these centered PSFs. 
\item Perform the standard pre-processing step using the deconvolved
  pinhole array images.
\item Do normal MOMFBD processing.
\item Deconvolve the restored images with the centered PSFs. 
\end{enumerate}

\section{Conclusions}
\label{sec:conclusions}

We have investigated a NB double-cavity filter, used at the SST for
scanning through the blue wing of the \ion{Ca}{ii}~H spectral line.

Measurements with the TRIPPEL spectrograph show that the filter has a
passband that is narrower by 10\% than indicated by profiles obtained
from the manufacturer. We note that the filter was 8 years old at the
time of these measurements but we know too little about coating age
effects to speculate on why this can happen. These measurements also
indicate that a description based on wavelength shifts of
single-cavity filter profiles, together with a model for variations in
the incident angle over the pupil, work very well for predicting the
position and shape of the broadened passband.

We calculate PSFs and OTFs corresponding to a diffraction limited
telescope and tilt angles in the range 0\fdg0--7\fdg5 and find that
the Strehl ratio decreases to less than 0.6 for the largest angle. 

We find that the tilt effects couple very weakly with wavefront
corrections and that the OTFs are separable in the usual OTF for phase
aberrations (including atmospheric turbulence effects) and \Stilt, a
correction for the filter tilt effects. This means it is sufficient to
correct images after restoration for atmospheric effects, the raw data
do not have to be corrected before image restoration. However, the
PSFs are in general off-center and asymmetric, which has consequences
for MOMFBD alignment. We have designed a procedure for \Stilt{}
compensation in a way that maintains the alignment properties of
MOMFBD image restoration.

By experiments with synthetic data, we find that deconvolution with the
proper OTF compensates very well for the degrading effects that are
really affecting the image formation at each wavelength in the
passband separately. We demonstrate the effect of the phase of the
transmission profile and show that not allowing for this phase in the
right way can significantly increase the RMS error in the compensated
images.

Our theoretical model for pupil apodisation effects from varying
incident angles on NB interference filters does not adequately predict
the effects on image quality. This is evident from power spectra of
pinhole images as well as solar data. This can be an error in our
derivations or coding, but we note that it would have to be a mistake
that affects \Stilt{} without causing errors in the profile
broadening.

Instead, we find that useful \Stilt{} can be measured by use of
pinhole calibration data. The ones we measure are visibly affected by
noise but this can easily be improved by use of data with better SNR.
Note that the better the solar data, the better pinhole images are
needed.

For the specific setup with a tilt-tunable \ion{Ca}{ii}~H filter at
the SST, the degrading effect on the PSF is negligible when the filter
is not tilted, but becomes more significant as the tilt angle is
increased.

A filter wheel with fixed passband filters or an FPI are better ways
of sampling different positions in a spectral line. However, tuning
interference filters by tilting them from normal incidence is useful
because it is a cheaper solution. It may be preferable if one can live
with the spectral broadening and compensate for the degraded
resolution.

\begin{acknowledgements}       
  
  G\"oran Scharmer initiated this research several years ago, early
  results were presented already by \citet{lofdahl04image}, and he has
  also provided many useful comments after the project was recently
  revived.
  Luc Rouppe van der Voort took part in the observational testing of
  the filter with TRIPPEL. Mats Carlsson is thanked for sharing the 3D
  MHD simulation snapshot used for making synthetic data.
  The Swedish 1-m Solar Telescope is operated on the island of La
  Palma by the Institute for Solar Physics of the Royal Swedish
  Academy of Sciences in the Spanish Observatorio del Roque de los
  Muchachos of the Instituto de Astrof\'isica de Canarias.


\end{acknowledgements}        
\balance


\appendix

\section{The incident angle}
\label{sec:incident-angle}

This appendix gives a derivation of an expression for the incident
angle $\alpha$ on a filter in a converging beam as a function of
position in the pupil. We use a coordinate system with the unit length
equal to the pupil diameter, $D$. The distance between the pupil and
focal planes is then equal to the F-ratio, $F=f/D$, where $f$ is the
focal length.

\begin{figure}[!t]
  \centering
  \subfloat[Pupil plane]{\includegraphics[bb=41 600 292 749,clip,width=\linewidth]{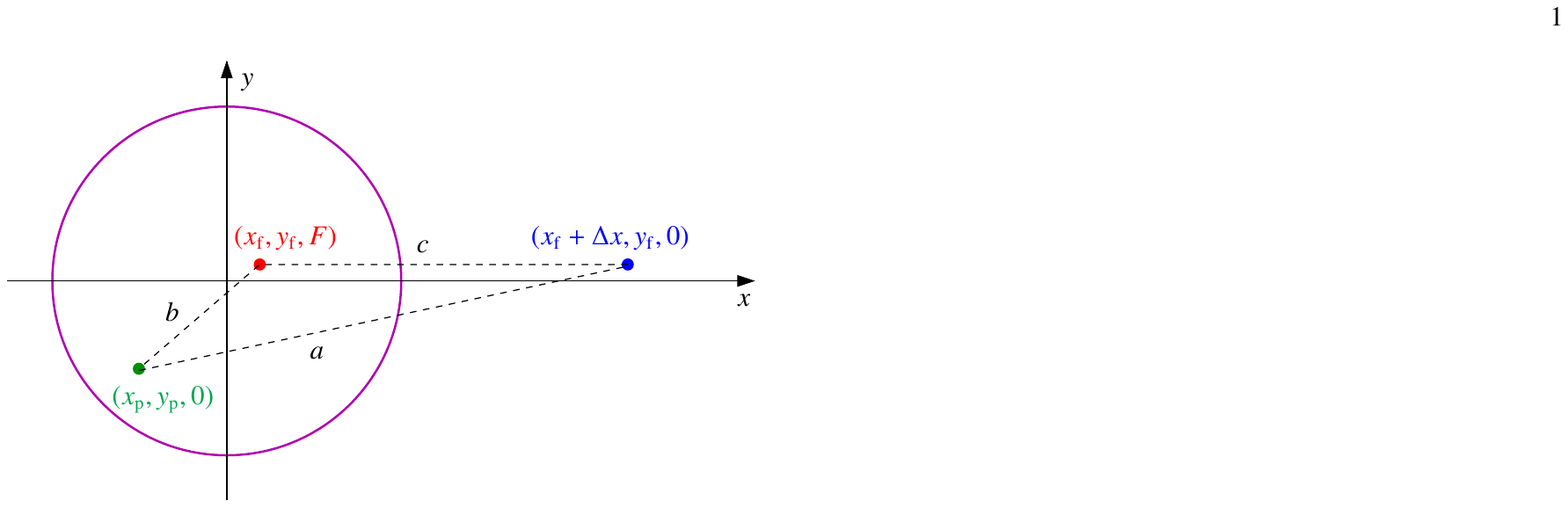}}\\
  \subfloat[Horizontal plane]{\includegraphics[bb=41 635 292 749,clip,width=\linewidth]{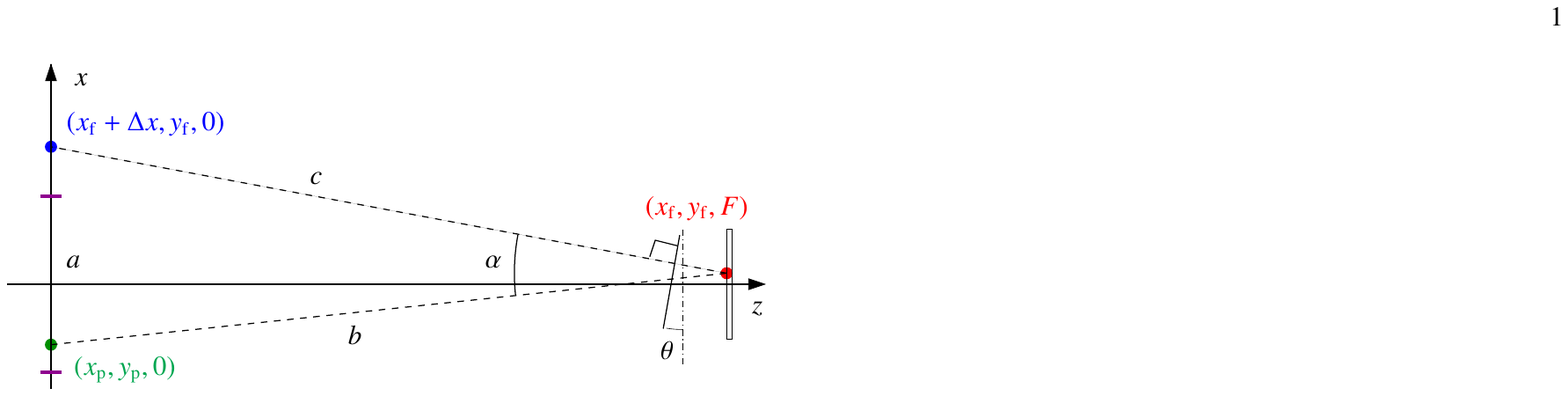}}
  \caption{Filter tilt geometry. \textbf{(a)} Projected on the pupil
    plane. \textbf{(b)} Top view, projection on the horizontal plane.
    The pupil with diameter $D=1$ is indicated by the circle in a) and
    by the two tick marks on the $x$ axis in b). The focal length is
    $f$. In $(x,y,z)$ space ($z$ is the optical axis), a ray leaves
    the pupil plane at $(\xp,\yp,0)$ and hits the detector in the
    focal plane at $(\xf,\yf,F)$. The filter surface normal from
    $(\xf,\yf,F)$ intersects with the pupil plane at an $x$-coordinate
    increased by an amount $\Delta x$ given by Eq.~(\ref{eq:xc}). The
    angle $\theta$ is in a horizontal plane but $\alpha$ is in general
    not.}
  \label{fig:tilt-geometry}
\end{figure}

A ray emanates from the pupil plane at the point $(\xp,\yp,0)$ and
intersects the detector in the focal plane at $(\xf,\yf,F)$.  The
filter can be anywhere between the pupil and the detector and it is
tilted by an angle $\theta$ around an axis parallel to the $y$
axis. The geometry of the pupil--filter system is shown in
Fig.~\ref{fig:tilt-geometry}.

The Law of Cosines gives a relation between the incident angle
$\alpha$ and the lengths of the sides in the $abc$ triangle,
\begin{equation}
  \label{eq:cos}
  a^2 = b^2 + c^2 - 2bc\cos\alpha,
\end{equation}
where $a$, $b$, and $c$ are the distances between the triangle
corners,
\begin{align}
  \label{eq:a}
  a^2 &= (\xf +\Delta x - \xp)^2+(\yf - \yp)^2,\\
  \label{eq:b}
  b^2 &= (\xp-\xf)^2 + (\yp-\yf)^2  + F^2,\\
  \label{eq:c}
  c^2 &= \Delta x^2 + F^2.
\end{align}
The length $\Delta x$ is given by the filter tilt angle $\theta$ as
\begin{equation}
  \label{eq:xc}
  \Delta x = F \tan\theta.
\end{equation}

For $\abs{\theta} \le \pi/2$, Eqs.~(\ref{eq:cos})--(\ref{eq:xc}) and
trivial algebra yield
\begin{equation}
  \label{eq:cosb}
  \cos\alpha =  
  \frac{F\cos\theta + (\xp-\xf)\sin\theta}
  {\bigl(F^2 + (\xp-\xf)^{2} + (\yp-\yf)^{2}\bigr)^{1/2}}.
\end{equation}

\end{document}